\documentclass[aps,prb,onecolumn,superscriptaddress,showpacs,amsmath,amssymb]{revtex4-2}

\usepackage{graphicx, comment}
\usepackage{dcolumn}
\usepackage{bm}
\usepackage{textcomp}
\usepackage{lineno}
\usepackage{chemformula}
\usepackage{multirow,makecell}
\usepackage{float}
\usepackage{ctable}

\begin{document}

%\preprint{}

\title{All-plasmonic switching effect in the graphene nanostructures containing quantum emitters}


\author{M.Yu. Gubin}
\affiliation{Department of Physics and Applied Mathematics, Vladimir State University named after Alexander and Nikolay Stoletovs (VlSU), Vladimir 600000, Russia}
\affiliation{Center for Photonics and 2D Materials, Moscow Institute of Physics and Technology (MIPT), Dolgoprudny 141701, Russia}
\author{A.Yu. Leksin}
\affiliation{Department of Physics and Applied Mathematics, Vladimir State University named after Alexander and Nikolay Stoletovs (VlSU), Vladimir 600000, Russia}
\author{A.V. Shesterikov}
\affiliation{Department of Physics and Applied Mathematics, Vladimir State University named after Alexander and Nikolay Stoletovs (VlSU), Vladimir 600000, Russia}
\affiliation{Center for Photonics and 2D Materials, Moscow Institute of Physics and Technology (MIPT), Dolgoprudny 141701, Russia}
\author{A.V. Prokhorov}
\email{alprokhorov33@gmail.com}
\affiliation{Department of Physics and Applied Mathematics, Vladimir State University named after Alexander and Nikolay Stoletovs (VlSU), Vladimir 600000, Russia}
\affiliation{Center for Photonics and 2D Materials, Moscow Institute of Physics and Technology (MIPT), Dolgoprudny 141701, Russia}
\author{V.S. Volkov}
\affiliation{Center for Photonics and 2D Materials, Moscow Institute of Physics and Technology (MIPT), Dolgoprudny 141701, Russia}
%\thanks{}
%\affiliation{}


%\date{\today}

\begin{abstract}
Nonlinear plasmonic effects in perspective 2D materials containing low-dimensional quantum emitters can be a basis of a novel technological platform for the fabrication of fast all-plasmonic triggers, transistors, and sensors. This article considers the conditions for achieving a strong plasmon-exciton coupling in the system with quantum nanowire (NW) placed in proximity to the nanostructured graphene sheets. In the condition of strong coupling, nonlinear interaction between two surface plasmon-polariton (SPP) modes propagating along the graphene waveguide integrated with a stub nanoresonator loaded with a core-shell semiconductor NWs is investigated. Using the 2D full-wave electromagnetic simulation, we studied the different transmittance regimes of the stub with NW for both the strong pump SPP and weak signal SPP tuned to interband and intraband transition in NW, respectively. We found such a regime of NW-SPP interaction that corresponds to the destructive interference with the signal SPP transmittance through the stub less than $7 \%$ in the case for pump SPP to be turned off. In contrast, the turning on the pump SPP leads to a transition to constructive interference in the stub and enhancement of signal SPP transmittance to $93 \%$. In our model, the effect of plasmonic switching occurs with a rate of $50 \; \textrm{GHz}$ at wavelength $8 \; \textrm{\textmu m}$ for signal SPP localized inside $20 \; \textrm{nm}$ graphene stub loaded with core-shell InAs/ZnS NW.
\end{abstract}

%\pacs{}

\maketitle

\section{Introduction}
The achievements of modern 2D material science~\cite{ali,Ponomarenko,nov2}, graphene nanotechnologies~\cite{ali,Guo1,Giubileo} and nanoplasmonics~\cite{boj1} give hope to the fabrication of novel ultra-fast plasmonic nanodevices in soon time. Such devices should be based on the new methods of surface plasmon-polariton (SPP) manipulations~\cite{Hosseininejad} in graphene, a good feature of which is the high localization of the electromagnetic field at the interface. The interaction between SPP modes in these plasmonic nanostructures can be realized through the use of both electronic and optical nonlinearities. Such nonlinearities can be achieved through the interaction of several graphene SPPs with chromophores (emitters) placed in the proximity of the graphene sheet. However, the efficiency of such an interaction strongly depends on the conditions of SPP-chromophore coupling. This means that the SPP-chromophore coupling constant should exceed the characteristic rate of electron scattering in graphene~\cite{Koppens} and the spontaneous relaxation rate in the chromophore~\cite{fedor,Prokhorov}. The last condition becomes very important since the spontaneous relaxation rate of the chromophore is strongly modified under the increase in the local density of optical states (LDOS) near the conductive surface.

This paper presents the results of analytical and numerical simulation of propagation the near- and mid-infrared electromagnetic fields localized on graphene sheets. The features of the SPP propagation through the empty graphene stub nanoresonator integrated with graphene waveguide are studied. It is shown that the tuning of the stub height leads to the reduction of the signal SPP transmittance through the waveguide almost to zero. We proposed to load a semiconductor nanowire (NW) into a stub nanoresonator and use it for the achievement of strong SPP-chromophore coupling using Ladder-type SPP-NW interaction scheme. Optimizing the NW parameters, we have shown the possibility to control the transmittance of the signal SPP by changing the intensity of the pump SPP. In particular, turning off the pump SPP, the transmittance of the signal SPP mode is kept constant at a level close to zero, but when the pump SPP is turned on, the signal SPP transmittance achieves $93\%$.

To accomplish this goal, we developed a quasiclassical approach to describe nonlinear plasmon-exciton interactions in multi-photon schemes~\cite{our3}, and also demonstrated the possibility for the realization of strong coupling conditions in a high-LDOS system. In addition, we used analytical and numerical methods to analyze the stability of the steady-state regimes of the waveguide transmittance. The presented approaches can be used for further development of the nonlinear theory of plasmon-exciton interactions in strong-coupling condition for a high-LDOS system. At the same time, the discussed applied effect of all-plasmonic switching may have a crucial role to play in the implementation of ultrafast plasmon transistors and systems with ``ultra-fast response'' based on them.

%%%%%%%%%%%%%%%%%%%%%%%%%%%%%%%%%%%%%%%%%%
\section{Mathematical models for the electrical conductivity of single graphene sheet and two coupled graphene sheets}

We start with the consideration of the propagation problem for the surface plasmon-polaritons in 2D graphene structures~\cite{sup1}. The electromagnetic field couples with the graphene sheet and then SPPs start to propagate along it~\cite{Koppens,Grigorenko} only if the photon energy is less than doubled chemical potential $\mu_{c}$ of graphene, $\hbar \omega <2\mu_{c}$~\cite{ali}. Because under this condition, the real part of dielectric permittivity becomes negative, i.e., graphene demonstrates metal-like properties. For example, the real part of permittivity becomes negative at wavelengths longer than $1.5 \; \textrm{\textmu m}$ for highly doped graphene with value $\mu_c=0.6 \; \textrm{eV}$ (further we assume that the electron scattering time $\tau$ is $0.9 \; \textrm{ps}$) taken from the literature~\cite{Jablan,teng}.

\begin{figure}[t]
\centering
\includegraphics[width=0.5\columnwidth]{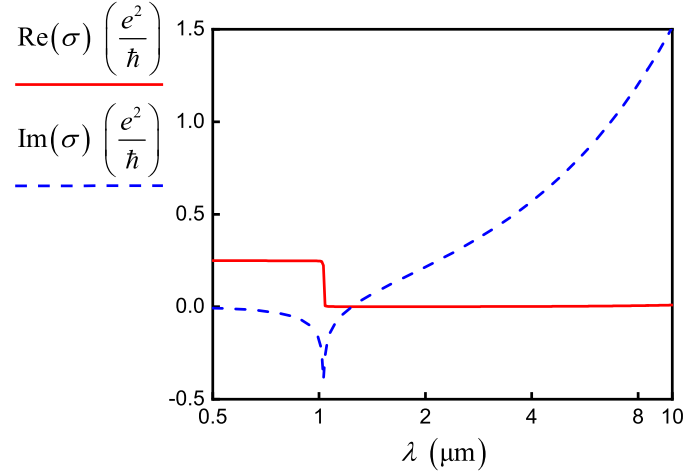}
\caption{\label{fig:1} Real (solid red curve) and imaginary (dashed blue curve) parts of the conductivity of doped graphene with $\mu_{c}=0.6 \; \textrm{eV}$, $\tau=0.9 \; \textrm{ps}$.}
\end{figure}

In general case, the total conductivity of graphene can be described by the Kubo formula~\cite{Mikhailov}:
\begin{align}
\label{eq:1}
\nonumber
\sigma\left(\omega,\mu_{c},\tau,T\right)=&\frac{-ie^2/\pi\hbar^2}{\omega +i/\tau } \int^{\infty}_{0}{\epsilon \left(\frac{\partial f_{d}\left(\epsilon \right)}{\partial \epsilon}-\frac{\partial f_{d}\left(-\epsilon \right)}{\partial \epsilon }\right)d\epsilon}\\
&-ie^2\left(\omega +i/\tau \right)/\pi \hbar^2 \int^{\infty}_{0}{\frac{f_{d}\left(\epsilon \right)-f_{d}\left(-\epsilon \right)}{{\left(\omega +i/\tau \right)}^2-4{\left(\epsilon /\hbar \right)}^2}d\epsilon},
\end{align}
where $1/\tau$ is the scattering rate of electrons, $f_{d}\left(\epsilon \right)=1/\left(e^{\left(\epsilon -\mu_{c}\right)/kT}+1\right)$ is the Fermi-Dirac distribution function, $T$ is the temperature (further we everywhere assume $T=300 \; \textrm{K}$), $k$ is the Boltzmann constant, $\hbar \equiv \frac{h}{2\pi}$, $h$ is the Planck constant, $e$ is the electron charge, see Fig.~\ref{fig:1}.

The expression (\ref{eq:1}) can be separated into two parts, one of which corresponds to the intraband conductivity approximated in the form
\begin{equation}
\label{eq:2}
\sigma_{\textrm{intra}}\left(\omega,\mu_{c},\tau,T\right)=i\frac{8\sigma_{0}kT/h}{\omega +i/\tau}\left(\frac{\mu_{c}}{kT}+2\textrm{ln}\left(e^{-\frac{\mu_{c}}{kT}}+1\right)\right),
\end{equation}
where $\sigma_{0}=\pi e^2/\left(2h\right)$. For the case when $kT \ll \left|\mu_{c}\right|,\hbar \omega $ the second integral in \ref{eq:1} can be approximated as follows
\begin{equation}
\label{eq:3}
\sigma_{\textrm{inter}}\left(\omega,\mu_{c},\tau,T\right)\approx i\frac{\sigma_{0}}{\pi} \textrm{ln}\left(\frac{2\mu_{c}-\left(\omega +i/\tau \right)\hbar}{2\mu_{c}+\left(\omega +i/\tau \right)\hbar}\right).
\end{equation}

\begin{figure}[t]
\centering
\includegraphics[width=0.5\columnwidth]{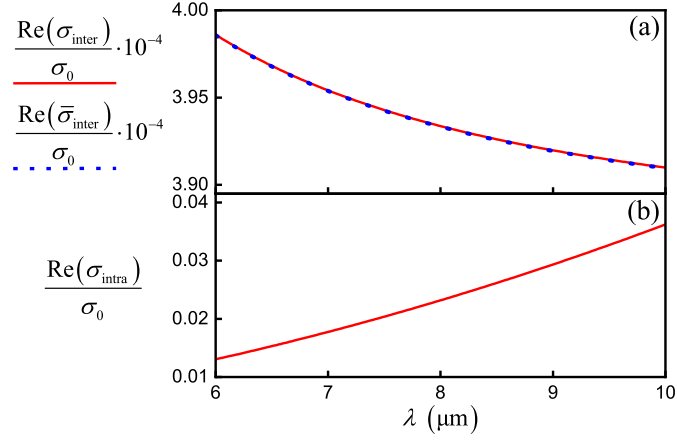}
\caption{\label{fig:2} (a) The dependence of interband conductivity $\sigma_{\textrm{inter}}$ (solid red line) and Pade approximation $\overline{\sigma}_{\textrm{inter}}$ (dotted blue line) normalized to $\sigma_{0}$ on the wavelength. (b) The dependence of intraband conductivity $\sigma_{\textrm{intra}}$ normalized to $\sigma_{0}$ on the wavelength, $\mu_{c}=0.6 \; \textrm{eV}$, $\tau =0.9 \; \textrm{ps}$.}
\end{figure}

Intraband conductivity becomes dominant under the condition $\mu_{c} > \hbar \omega$, as well as the interband conductivity, takes considerable values under condition $\mu_{c} < \hbar \omega$. Thus, for terahertz, far- and mid-infrared radiations and $\mu_{c}=0.6 \; \textrm{eV}$, the effect of interband conductivity can be neglected~\cite{ali}. This is confirmed by the dependence of inter- and intraband conductivity for graphene on wavelength shown in Figs.~\ref{fig:2}a and~\ref{fig:2}b.

The interband conductivity (\ref{eq:3}) can be approximated by the Pade formula~\cite{pade} in the form:
\begin{equation}
\label{eq:4}
\overline{\sigma}_{\textrm{inter}}\left(\omega\right)=\frac{a_{0}+a_{1}\cdot \left(i\omega \right)+a_{2}\cdot {\left(i\omega \right)}^2}{1+b_{1}\cdot \left(i\omega \right)+b_{2}\cdot {\left(i\omega \right)}^2}.
\end{equation}
The corresponding coefficients can be found by solving the system of equations:
\begin{equation}
\label{eq:5}
\begin{bmatrix}
1 & 0 & -\omega_{p1}^{2} & \omega_{p1}\Theta\left(\omega_{p1}\right) & \omega_{p1}^{2}\Gamma\left(\omega_{p1}\right)\\
0 & \omega_{p1} & 0 & -\omega_{p1}\Gamma\left(\omega_{p1}\right) & \omega_{p1}^{2}\Theta\left(\omega_{p1}\right)\\
1 & 0 & -\omega_{p2}^{2} & \omega_{p2}\Theta\left(\omega_{p2}\right) & \omega_{p2}^{2}\Gamma\left(\omega_{p2}\right)\\
0 & \omega_{p2} & 0 & -\omega_{p2}\Gamma\left(\omega_{p2}\right) & \omega_{p2}^{2}\Theta\left(\omega_{p2}\right)\\
1 & 0 & -\omega_{p3}^{2} & \omega_{p3}\Theta\left(\omega_{p3}\right) & \omega_{p3}^{2}\Gamma\left(\omega_{p3}\right)
\end{bmatrix}
\begin{pmatrix}
a_{0}\\ a_{1}\\ a_{2}\\ b_{1}\\ b_{2}
\end{pmatrix}=
\begin{pmatrix}
\Gamma\left(\omega_{p1}\right)\\ \Theta\left(\omega_{p1}\right)\\ \Gamma\left(\omega_{p2}\right)\\ \Theta\left(\omega_{p2}\right)\\ \Gamma\left(\omega_{p3}\right)
\end{pmatrix},
\end{equation}
where $\Theta\left(\omega \right)=\textrm{Im}\left(\sigma_{\textrm{inter}}\right)$ and $\Gamma\left(\omega \right)=\textrm{Re}\left(\sigma_{\textrm{inter}}\right)$. In particular, using parameters from Table~\ref{tab:1} (for approximation nearby $\lambda =8 \; \textrm{\textmu m}$ we obtained the following values of coefficients $a_{0}=2.346\cdot {10}^{-8}$, $a_{1}=-2.112\cdot {10}^{-20}$, $a_{2}=9.589\cdot {10}^{-39}$, $b_{1}=-6.745\cdot {10}^{-19}$, $b_{2}=1.007\cdot {10}^{-31}$ within the fitting of Kubo formula by the following three reference wavelengths: $\lambda_{p1}=7.2 \; \textrm{\textmu m}$, $\lambda_{p2}=8.2 \; \textrm{\textmu m}$, $\lambda_{p3}=9.2 \; \textrm{\textmu m}$ ($\omega_{pi}=\frac{2\pi c}{\lambda_{pi}}$, $i=1,2,3$). In this case, the dielectric permittivity of graphene sheet with the effective thickness $\Delta_{g}$ can be calculated as follows
\begin{equation}
\label{eq:6}
\varepsilon_{gr}=1+i\frac{\sigma_{\textrm{intra}}}{\omega \Delta_{g}\varepsilon_{0}}=1+i\frac{\sigma_{1}}{\omega \varepsilon_{0}\left(1-i\omega \tau \right)},
\end{equation}
where a new parameter $\sigma_{1}=\frac{e^2kT\tau}{\pi \hbar^{2}\Delta_{g}}\left(\frac{\mu_{c}}{kT}+2\textrm{ln}\left(e^{-\frac{\mu_{c}}{kT}}+1\right)\right)$ was introduced~\cite{Sarker}. Here, it should be noted that for the numerical algorithms, we use the relation $\sigma_{\textrm{intra}}=\frac{\sigma_{1}\Delta_{g}}{1-i\omega \tau}$ and the effective thickness $\Delta_{g}$. For the realization of the 2D finite difference time domain (FDTD) method, the permittivity of graphene is represented in the following form~\cite{Sarker}:
\begin{equation}
\label{eq:7}
\varepsilon_{gr}=1+i\frac{\sigma_{1}}{\omega \varepsilon_{0}}-\frac{\tau \sigma_{1}}{\left(1-i\omega \tau \right)\varepsilon_{0}}.
\end{equation}

\begin{table}[H]
\caption{The characteristics of SPP generated at the graphene sheets with parameters: $\mu_{c}=0.6 \; \textrm{eV}$, $\tau=0.9 \; \textrm{ps}$, $T=300 \; \textrm{K}$, $\Delta_{g}=2 \; \textrm{nm}$, $d=20 \; \textrm{nm}$.}
\label{tab:1}
\centering
\footnotesize
\begin{tabular}{*{9}{c}}
\toprule
$\lambda_{0}, \; \textrm{\textmu m}$ & $\varepsilon_{d}$ & $\frac{2\mu_{c}}{\hbar \omega}$ & \multicolumn{2}{c}{$\sigma_{1}, \; \textrm{S}/\textrm{m}$} & \multicolumn{2}{c}{$\sigma_{\textrm{intra}}, \; \textrm{S}$} & \multicolumn{2}{c}{$\sigma_{\textrm{inter}}, \; \textrm{S}$} \\
\midrule
\multirow{2}{*}{$4$} &  $1$ (air) & $3.88$ & \multicolumn{2}{c}{$3.193\cdot 10^{7}$} & \multicolumn{2}{c}{$3.5\cdot 10^{-7}+1.49\cdot 10^{-4}i$} & \multicolumn{2}{c}{$2.51\cdot 10^{-8}-1.02\cdot 10^{-5}i$} \\
& $2.103$ (\ch{SiO2}) & $3.88$ & \multicolumn{2}{c}{$3.193\cdot 10^{7}$} & \multicolumn{2}{c}{$3.5\cdot 10^{-7}+1.49\cdot 10^{-4}i$} & \multicolumn{2}{c}{$2.51\cdot 10^{-8}-1.02\cdot 10^{-5}i$} \\
$1.96$ & $2.103$ (\ch{SiO2}) & $1.9$ & \multicolumn{2}{c}{$3.193\cdot 10^{7}$} & \multicolumn{2}{c}{$8.4\cdot 10^{-8}+7.3\cdot 10^{-5}i$} & \multicolumn{2}{c}{$3.24\cdot 10^{-8}-2.26\cdot 10^{-5}i$} \\
$2.56$ & $2.022$ & $2.483$ & \multicolumn{2}{c}{$3.193\cdot 10^{7}$} & \multicolumn{2}{c}{$1.44\cdot 10^{-7}+9.54\cdot 10^{-5}i$} & \multicolumn{2}{c}{$2.8\cdot 10^{-8}-1.65\cdot 10^{-5}i$} \\
$8.04$ & $2.022$ & $7.8$ & \multicolumn{2}{c}{$3.193\cdot 10^{7}$} & \multicolumn{2}{c}{$1.42\cdot 10^{-6}+3\cdot 10^{-4}i$} & \multicolumn{2}{c}{$2.39\cdot 10^{-8}-4.98\cdot 10^{-6}i$} \\
\bottomrule \\
%& & & & & & & & \\
\toprule
\multirow{2}{*}{$\lambda_{0}, \; \textrm{\textmu m}$} & \multirow{2}{*}{$\varepsilon_{d}$} & \multicolumn{2}{|c|}{single layer} & \multicolumn{5}{c}{double-layer sheet} \\
& & \multicolumn{1}{|c}{$\lambda_{SPP}, \; \textrm{nm}$} & \multicolumn{1}{c|}{$L_{SPP}, \; \textrm{\textmu m}$} & $\textrm{Re}\left(\xi \right), \; \textrm{nm}$ & $n^{\left(\textrm{R}\right)}_{EF+}$ & $\lambda_{SPP+}, \; \textrm{nm}$ & $L_C, \; \textrm{nm}$ & $\overline{L}_{SPP+}, \; \textrm{\textmu m}$ \\
\midrule
\multirow{2}{*}{$4$} &  $1$ (air) & \multicolumn{1}{|c}{$104.6$} & \multicolumn{1}{c|}{$3.1$} & $33$ & $49$ & $81.5$ & $61$ & $3.5$ \\
& $2.103$ (\ch{SiO2}) & \multicolumn{1}{|c}{$49.7$} & \multicolumn{1}{c|}{$1.5$} & $15.8$ & $86.1$ & $46.5$ & $108.3$ & $1.6$ \\
$1.96$ & $2.103$ (\ch{SiO2}) & \multicolumn{1}{|c}{$8.86$} & \multicolumn{1}{c|}{$0.3$} & $2.82$ & $221$ & $8.86$ & $2.3\cdot 10^{6}$ & $0.3$ \\
$2.56$ & $2.022$ & \multicolumn{1}{|c}{$18.8$} & \multicolumn{1}{c|}{$0.7$} & $6$ & $136$ & $18.8$ & $2641$ & $0.7$ \\
$8.04$ & $2.022$ & \multicolumn{1}{|c}{$224.2$} & \multicolumn{1}{c|}{$3.8$} & $71$ & $59.3$ & $135.5$ & $74$ & $3.7$ \\
\bottomrule
\end{tabular}
\end{table}

The wave vector of SPP propagating along a single graphene sheet can be written as follows
\begin{equation}
\label{eq:8}
k_{SPP}=k_{0}\sqrt{\varepsilon_{d}-{\left(\frac{2\varepsilon_{d}\varepsilon_{0}c}{\sigma_{g}}\right)}^2},
\end{equation}
where $\sigma_{g}\equiv \sigma_{\textrm{intra}}$, $k_{0}=\frac{2\pi}{\lambda_{0}}$ is the wave vector of the electromagnetic field at a wavelength $\lambda_{0}$ in vacuum, $\varepsilon_{d}$ is the dielectric permittivity of the host medium, $\varepsilon_{0}$ is the electric constant, $c$ is the speed of light in vacuum. The wavelength of SPP localized on the graphene sheet has the form $\lambda_{SPP}=\frac{2\pi}{k_{SPP}}$ and the propagation length (i.e., the characteristic distance of SPP decay) is given by the expression
\begin{equation}
\label{eq:9}
L_{SPP}=\frac{\lambda_{0}}{4\pi \textrm{Im}\left(\frac{k_{SPP}}{k_{0}}\right)}.
\end{equation}

Now we will consider the formation of coupled SPPs propagating along the two parallel graphene sheets placed at a small distance $d$ between them~\cite{ali,Hossain}. In this case, the dispersion relation for SPP propagation constants $\beta $ can be written in the form~\cite{teng}
\begin{equation}
\label{eq:10}
-k_{h}\left(\pm e^{-k_{h}d}-1\right)=2ik_{0}c\varepsilon_{d}\varepsilon_{0}/\sigma_{g},
\end{equation}
where $k_{h}=\sqrt{\beta^{2}-k^{2}_{0}}$. The solution $\beta_{+}$ corresponds to symmetric and $\beta_{-}$ corresponds to anti-symmetric SPP mode. We will discuss the symmetric mode only because it leads to the highest density of the electromagnetic field in the space between sheets. It is necessary to increase the efficiency of matter-field interaction with chromophore loaded in the space between sheets.

\begin{figure}[t]
\centering
\includegraphics[width=0.5\columnwidth]{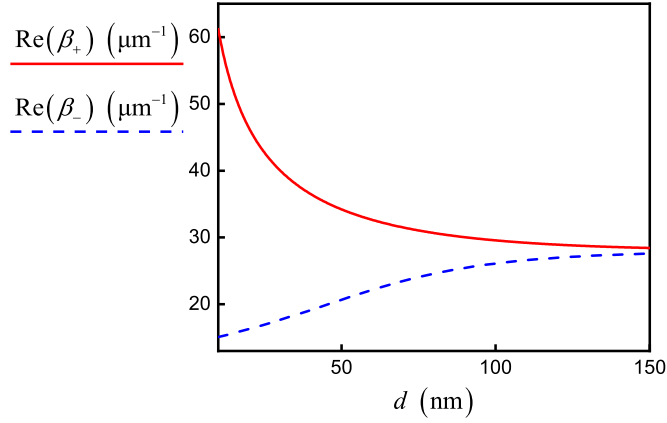}
\caption{\label{fig:3} Propagation constants $\beta_{+}$ (solid red line) and $\beta_{-}$ (dashed blue line) for signal SPP modes for the double-layer graphene sheets versus the interlayer distance $d$ numerically calculated from Equation (\ref{eq:10}). Parameters correspond to the strong coupling regime for wavelength $8.04 \; \textrm{\textmu m}$ from Table~\ref{tab:1}.}
\end{figure}

The strong and weak coupling can be realized between sheets. To determine the type of coupling, we compare the distance $d$ between sheets with the characteristic parameter $\xi$ given by
\begin{equation}
\label{eq:11}
\xi =\textrm{Re}\left(\frac{\sigma_{g}}{ic\varepsilon_{0}\varepsilon_{d}k_{0}}\right).
\end{equation}
In condition $d>\xi$, the dispersion curves have a hyperbolic form, and wave vectors of graphene plasmons coincide with the same ones for the case of single layer graphene, which corresponds to the weak coupling. The condition $d<\xi$ corresponds to the strong SPP-graphene coupling, and the dispersion curves can significantly differ from the same ones for a single sheet of graphene.

In our work, we use both the direct numerical simulation of (\ref{eq:10}) and its approximate analytical solution for the weak SPP-graphene coupling regime. In the last case, the propagation constants for the symmetric and antisymmetric SPP modes~\cite{teng} are given by the expressions $\beta_{+}=k_{SPP}+\Delta \beta_{+}$ and $\beta_{-}=k_{SPP}+\Delta \beta_{-}$, where $\Delta \beta_{+}$ and $\Delta \beta_{-}$ are the small quantities relative to $k_{SPP}$. After solving (\ref{eq:10}) the approximate expressions for $\beta_{\pm}$ have the forms:
\begin{equation}
\label{eq:12}
\beta_{\pm}\approx k_{SPP}+\frac{2i\varepsilon_{0}\varepsilon_{d}\omega_{2}/\sigma_{g}-k_{p}\left(1\mp u_{p}\right)}{\left(1\mp u_{p}\right)k_{SPP}/k_{p}\pm u_{p}k_{SPP}d},
\end{equation}
where $k_{p}=\sqrt{k^{2}_{SPP}-\varepsilon_{d}k^{2}_{0}}$ and $u_{p}=e^{-k_{p}d}$.

The values of propagation constants correspond to the formation of SPPs in graphene at the wavelengths $\lambda_{SPP\pm}=\frac{2\pi}{\textrm{Re}\left(\beta_{\pm}\right)}$ depending on distance $d$ between sheets. In this case, the effective refractive index can be determined as $n_{EF\pm}=n^{\left(\textrm{R}\right)}_{EF\pm}+in^{\left(\textrm{I}\right)}_{EF\pm}=\frac{\beta_{\pm}}{k_{0}}$ and the characteristic length of the coupling is given by the relation
\begin{equation}
\label{eq:13}
L_{C}=\frac{\pi}{2\sqrt{2}\left|C_{g}\right|},
\end{equation}
where $C_{g}$ is the coupling constant and it can be presented as $C_{g}=\frac{\beta_{-}-\beta_{+}}{2}$. The propagation length of SPP for two sheets is defined as $\overline{L}_{SPP\pm}=\frac{\lambda_{0}}{4\pi \textrm{Im}\left(n_{EF\pm}\right)}$.

\begin{figure}[t]
\centering
\includegraphics[width=\columnwidth]{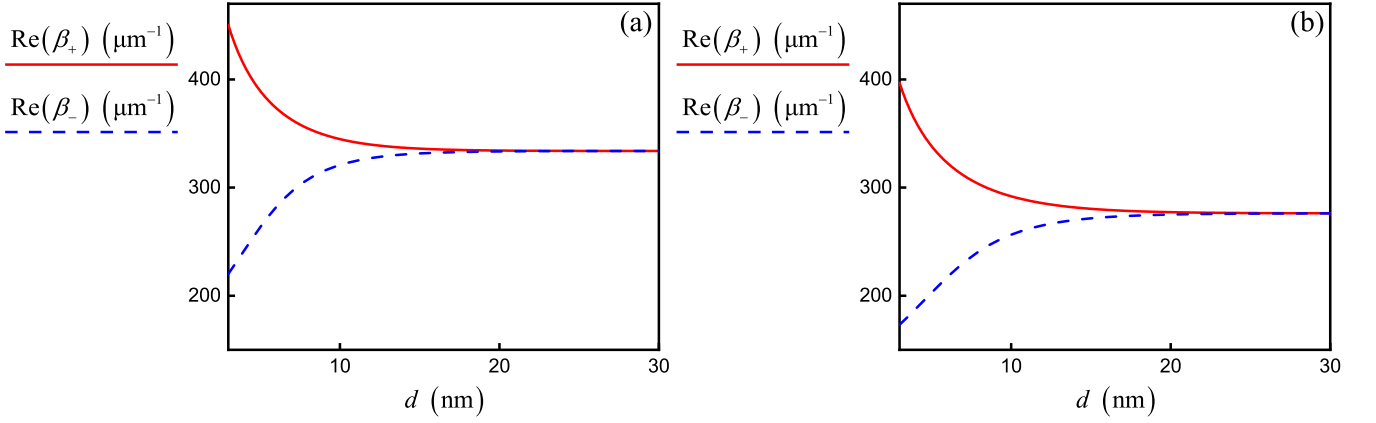}
\caption{\label{fig:4} Propagation constants $\beta_{+}$ (solid red lines) and $\beta_{-}$ (dashed blue lines) for SPP modes in the double-layer graphene sheets versus the interlayer distance $d$ (a) with and (b) without taking into account the interband conductivity of graphene calculated by using analytical solutions (\ref{eq:12}). Parameters correspond to a weak coupling regime for wavelength $2.56 \; \textrm{\textmu m}$ from Table~\ref{tab:1}.}
\end{figure}

Based on the simulation parameters from Table~\ref{tab:1} and fixed value $d=20 \; \textrm{nm}$, we obtained different regimes of coupling. For example, the initialization of SPP by electromagnetic field source with wavelength $8.04 \; \textrm{\textmu m}$ leads to the formation of a strong coupling regime with propagation constants that can be calculated only numerically by solving Equation (\ref{eq:10}), see Fig.~\ref{fig:3}. Note that interband conductivity does not influence the curves in Fig.~\ref{fig:3}. On the other hand, Fig.~\ref{fig:4} shows the curves calculated in accordance with (\ref{eq:12}) in the regime of weak SPP-graphene coupling for wavelength $2.56 \; \textrm{\textmu m}$. They are almost identical with the numerical solution of (\ref{eq:10}), but the contribution of interband conductivity increases at this wavelength. However, we will not take into account the correction associated with interband conductivity in FDTD simulation (see (\ref{eq:3})), which will slightly reduce the accuracy of our numerical experiments for $2.56 \; \textrm{\textmu m}$ wavelength.

\section{Numerical simulation of SPP generation in graphene sheets using the FDTD method}
\begin{figure}[t]
\centering
\includegraphics[width=0.4\columnwidth]{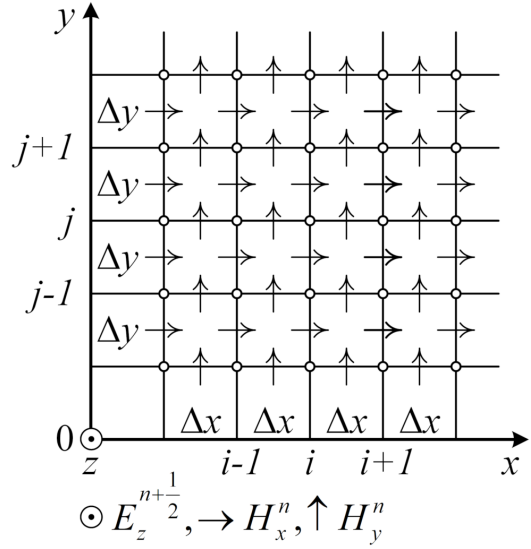}
\caption{\label{fig:5} The schematic illustration of the functioning numerical algorithm in the FDTD method for evaluating the graphene sheet.}
\end{figure}

We assume that the graphene sheet is located in plane $y=0$ in Fig.~\ref{fig:5}, and the source is the electric or magnetic dipole localized near the surface. In the two-dimensional case, all the functions do not change across $z$ axis, and the derivatives of these functions with respect to $z$ are zero. Then, the system splits into two parts corresponding to the TE and TM modes.

In this case, the evolution of the electromagnetic signal is described by two independent systems of equations for the components of electric field $E$, magnetic field $H$, and electric displacement $D$ in the form~\cite{Sullivan}:
$$
\begin{array}{|c|c|}
\hline
\textrm{TM-mode} & \textrm{TE-mode} \\
\hline
\frac{\partial D_{z}}{\partial t}=\frac{1}{\sqrt{\varepsilon_{0}\mu_{0}}}\left(\frac{\partial \widetilde{H}_y}{\partial x}-\frac{\partial \widetilde{H}_{x}}{\partial y}\right) & \frac{\partial D_{x}}{\partial t}=\frac{1}{\sqrt{\varepsilon_{0}\mu_{0}}}\frac{\partial \widetilde{H}_{z}}{\partial y} \\
\frac{\partial \widetilde{H}_{x}}{\partial t}=-\frac{1}{\sqrt{\varepsilon_{0}\mu_{0}}}\frac{\partial E_{z}}{\partial y} & \frac{\partial D_{y}}{\partial t}=-\frac{1}{\sqrt{\varepsilon_{0}\mu_{0}}}\frac{\partial \widetilde{H}_{z}}{\partial x} \\
\frac{\partial \widetilde{H}_{y}}{\partial t}=\frac{1}{\sqrt{\varepsilon_{0}\mu_{0}}}\frac{\partial E_{z}}{\partial x} & \frac{\partial \widetilde{H}_{z}}{\partial t}=\frac{1}{\sqrt{\varepsilon_{0}\mu_{0}}}\left(\frac{\partial E_{x}}{\partial y}-\frac{\partial E_{y}}{\partial x}\right) \\
\hline
\end{array}
$$
The quantities $E$ and $D$ are normalized:
$$E=\sqrt{\frac{\varepsilon_{0}}{\mu_{0}}}\widetilde{E}, \; D=\frac{1}{\sqrt{\varepsilon_{0}\mu_{0}}}\widetilde{D}.$$
The derivation of auxiliary difference equations for each mode was carried out using the PML method in the frequency domain. The following definitions were used:
$$\frac{1}{\sqrt{\varepsilon_{0}\mu_{0}}}=c, \; \frac{\partial }{\partial t}\rightarrow i\omega, \; i=\sqrt{-1},$$
where $c$ is the speed of light in vacuum. We obtained:
$$
\begin{array}{|c|c|}
\hline
\textrm{TM-mode} & \textrm{TE-mode} \\
\hline
i\omega D_{z}\varepsilon^{*}_{F_{z}}\left(x\right)\varepsilon^{*}_{F_{z}}\left(y\right)=C_{0}\left(\frac{\partial H_{y}}{\partial x}-\frac{\partial H_{x}}{\partial y}\right) & i\omega H_{z}\mu^{*}_{F_{z}}\left(x\right)\mu^{*}_{F_{z}}\left(y\right)=C_{0}\left(\frac{\partial E_{x}}{\partial y}-\frac{\partial E_{y}}{\partial x}\right) \\
i\omega H_{x}\mu^{*}_{F_{x}}\left(x\right)\mu^{*}_{F_{x}}\left(y\right)=-C_{0}\frac{\partial E_{z}}{\partial y} & i\omega D_{x}\varepsilon^{*}_{F_{x}}\left(x\right)\varepsilon^{*}_{F_{x}}\left(y\right)=C_{0}\frac{\partial H_{z}}{\partial y} \\
i\omega H_{y}\mu^{*}_{F_{y}}\left(x\right)\mu^{*}_{F_{y}}\left(y\right)=C_{0}\frac{\partial E_{z}}{\partial x} & i\omega D_{y}\varepsilon^{*}_{F_{y}}\left(x\right)\varepsilon^{*}_{F_{y}}\left(y\right)=-C_{0}\frac{\partial H_{z}}{\partial x} \\
\hline
\end{array}
$$
where permittivities of graphene are rewritten via dependencies of its conductivity on coordinates $\sigma_{g}\left(u\right)$ ($u=x,y$) as follows:
$$
\begin{array}{ll}
\varepsilon^{*}_{F_{z}}\left(x\right)=1+\frac{\sigma_{g}\left(x\right)}{i\omega \varepsilon_{0}} & \varepsilon^{*}_{F_{z}}\left(y\right)=1+\frac{\sigma_{g}\left(y\right)}{i\omega \varepsilon_{0}} \\
\mu^{*}_{F_{x}}\left(x\right)=\left(1+\frac{\sigma_{g}\left(x\right)}{i\omega \varepsilon_{0}}\right)^{-1} & \mu^{*}_{F_{x}}\left(y\right)=1+\frac{\sigma_{g}\left(y\right)}{i\omega \varepsilon_{0}} \\
\mu^{*}_{F_{y}}\left(x\right)=1+\frac{\sigma_{g}\left(x\right)}{i\omega \varepsilon_{0}} & \mu^{*}_{F_{y}}\left(y\right)=\left(1+\frac{\sigma_{g}\left(y\right)}{i\omega \varepsilon_{0}}\right)^{-1} \\
\mu^{*}_{F_{z}}\left(x\right)=1+\frac{\sigma_{g}\left(x\right)}{i\omega \varepsilon_{0}} & \mu^{*}_{F_{z}}\left(y\right)=1+\frac{\sigma_{g}\left(y\right)}{i\omega \varepsilon_{0}} \\
\varepsilon^{*}_{F_{x}}\left(x\right)=1+\frac{\sigma_{g}\left(x\right)}{i\omega \varepsilon_{0}} & \varepsilon^{*}_{F_{x}}\left(y\right)=1+\frac{\sigma_{g}\left(y\right)}{i\omega \varepsilon_{0}} \\
\varepsilon^{*}_{F_{y}}\left(x\right)=1+\frac{\sigma_{g}\left(x\right)}{i\omega \varepsilon_{0}} & \varepsilon^{*}_{F_{y}}\left(y\right)=1+\frac{\sigma_{g}\left(y\right)}{i\omega \varepsilon_{0}}
\end{array}
$$
For example, for considered TM-mode, we reordered the equations and obtained:
\begin{subequations}
\label{eq:14}
\begin{align}
i\omega \left(1+\frac{\sigma_{g}\left(x\right)}{i\omega \varepsilon_{0}}\right)^{-1}\left(1+\frac{\sigma_{g}\left(y\right)}{i\omega \varepsilon_{0}}\right)H_{x}&=-C_{0}\frac{\partial E_{z}}{\partial y}, \\
i\omega \left(1+\frac{\sigma_{g}\left(x\right)}{i\omega \varepsilon_{0}}\right)\left(1+\frac{\sigma_{g}\left(y\right)}{i\omega \varepsilon_{0}}\right)^{-1}H_{y}&=C_{0}\frac{\partial E_{z}}{\partial x}, \\
i\omega \left(1+\frac{\sigma_{g}\left(x\right)}{i\omega \varepsilon_{0}}\right)\left(1+\frac{\sigma_{g}\left(y\right)}{i\omega \varepsilon_{0}}\right)D_{z}&=C_{0}\left(\frac{\partial H_{y}}{\partial x}-\frac{\partial H_{x}}{\partial y}\right),
\end{align}
\end{subequations}
From equation (\ref{eq:14}a), we derived:
$$\omega \left(1+\frac{\sigma_{g}\left(y\right)}{i\omega \varepsilon_{0}}\right)H_{x}=-C_{0}\left(\frac{\partial E_{z}}{\partial y}+\frac{\sigma_{g}\left(x\right)}{i\omega \varepsilon_{0}}\frac{\partial E_{z}}{\partial y}\right)$$
or in another form:
\begin{equation}
\label{eq:15}
\frac{\partial H_{x}}{\partial t}+\frac{\sigma_{g}\left(y\right)}{\varepsilon_{0}}H_{x}=-C_{0}\left(\frac{\partial E_{z}}{\partial y}+\frac{\sigma_{g}\left(x\right)}{\varepsilon_{0}}\int^{T}_{0}\frac{\partial E_{z}}{\partial y}\partial t\right).
\end{equation}

Next, we use standard approximation:
\begin{align}
\nonumber
H_{x}&\approx \frac{H^{n+1}_{x}\left(i,j+\frac{1}{2}\right)+H^{n}_{x}\left(i,j+\frac{1}{2}\right)}{2}, \\
\nonumber
\frac{\partial H_{x}}{\partial t}&\approx \frac{H^{n+1}_{x}\left(i,j+\frac{1}{2}\right)-H^{n}_{x}\left(i,j+\frac{1}{2}\right)}{\Delta t}, \\
\nonumber
\int^{T}_{0}\frac{\partial E_{z}}{\partial y}\partial t&\approx \Delta t\sum^{n}_{k=0}{\frac{E^{n+\frac{1}{2}}_{z}\left(i,j+1\right)-E^{n+\frac{1}{2}}_{z}\left(i,j\right)}{\Delta x}}.
\end{align}
We introduce the definition:
$$curl\_x\left(n\right)=E^{n+\frac{1}{2}}_{z}\left(i,j\right)-E^{n+\frac{1}{2}}_{z}\left(i,j+1\right)$$
and from Equation (\ref{eq:15}) we obtain:
\begin{align}
\nonumber
&\frac{H^{n+1}_{x}\left(i,j+\frac{1}{2}\right)-H^{n}_{x}\left(i,j+\frac{1}{2}\right)}{\Delta t} \\
\nonumber
&+\frac{\sigma_{g}\left(j+\frac{1}{2}\right)}{\varepsilon_{0}}\frac{H^{n+1}_{x}\left(i,j+\frac{1}{2}\right)
+H^{n}_{x}\left(i,j+\frac{1}{2}\right)}{2}=C_{0}\left(\frac{curl\_x\left(n\right)}{\Delta x}+\frac{\sigma_{g}\left(i\right)\Delta t}{\varepsilon_{0}\Delta x}\sum^{n}_{k=0}{curl\_x\left(k\right)}\right),
\end{align}
where $i,j$ are the spatial coordinate indexes, $n$ is the time coordinate index, $\Delta x$ is the step along the spatial axis, $\Delta t$ is the step along the time axis.

A similar transformation for other Equations (\ref{eq:14}) was carried out, and we obtained the self-consistent system of equations for all field components and numerically realized the algorithm of calculation of these components. At the same time, the field source is a harmonic function in the form
$$D_{z}\left(i,j\right)=\textrm{sin}\left(\omega_{0}t\right),$$
where $\omega_{0}$ ($\lambda_{0}=\frac{2\pi c}{\omega_{0}}$) is the frequency (wavelength) of electromagnetic field source.

\begin{figure}[t]
\centering
\includegraphics[width=0.5\columnwidth]{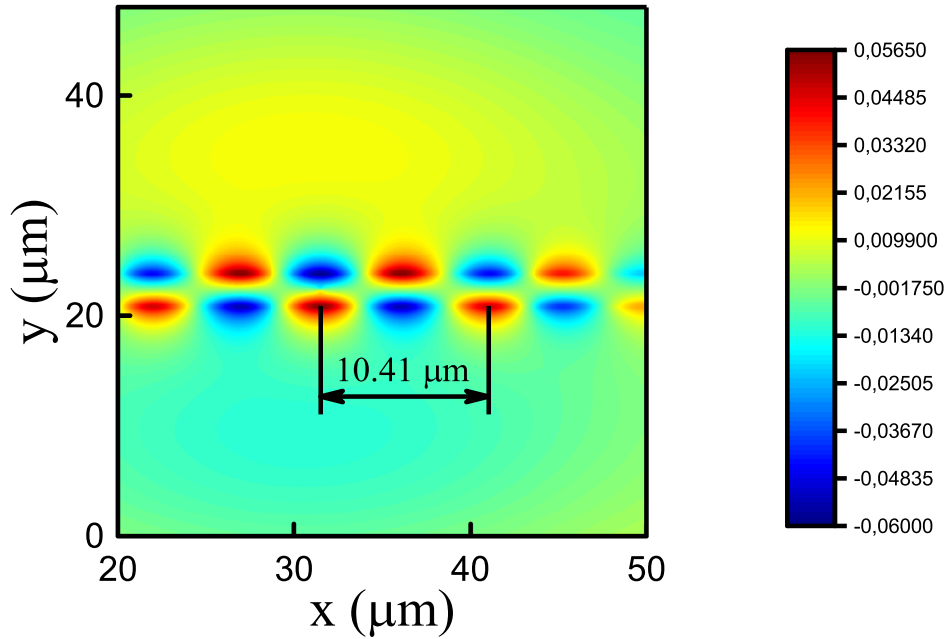}
\caption{\label{fig:6} The spatial distribution of field component $E_{y}$ for SPP generated on a pair of graphene sheets. The FDTD method is our own implementation in MATLAB. The parameters correspond to~\cite{Hossain}. Taking into account new values of conductivities~\cite{Hossain} $\sigma_{1}=8.91\cdot 10^{4} \; \textrm{S}/\textrm{m}$, $\sigma_{\textrm{intra}}=9.7\cdot 10^{-5}+1.6\cdot 10^{-3}i \; \textrm{S}$ (slight differences from~\cite{Hossain} are associated with calculation accuracy). The calculated value $\lambda_{SPP+}=10.41 \; \textrm{\textmu m}$.}
\end{figure}

We numerically realized the FDTD algorithm and developed an application for calculating electromagnetic modes in the proximity of graphene sheets. Comparing our simulation results for single and double sheets of graphene with known results, we concluded that our FDTD realization is in good agreement with them, see~\cite{teng,Hossain} and Fig.~\ref{fig:6}. Then, using our application, we performed original full-wave electromagnetic simulation for graphene sheets with different wavelengths of source, see Table~\ref{tab:1}. All numerical results correspond to analytical estimations in accordance to (\ref{eq:9})--(\ref{eq:13}).

\section{The model of Ladder-type nonlinear interactions between two SPPs and semiconductor NW loaded into graphene stub nanoresonator}
\begin{figure}[t]
\centering
\includegraphics[width=0.7\columnwidth]{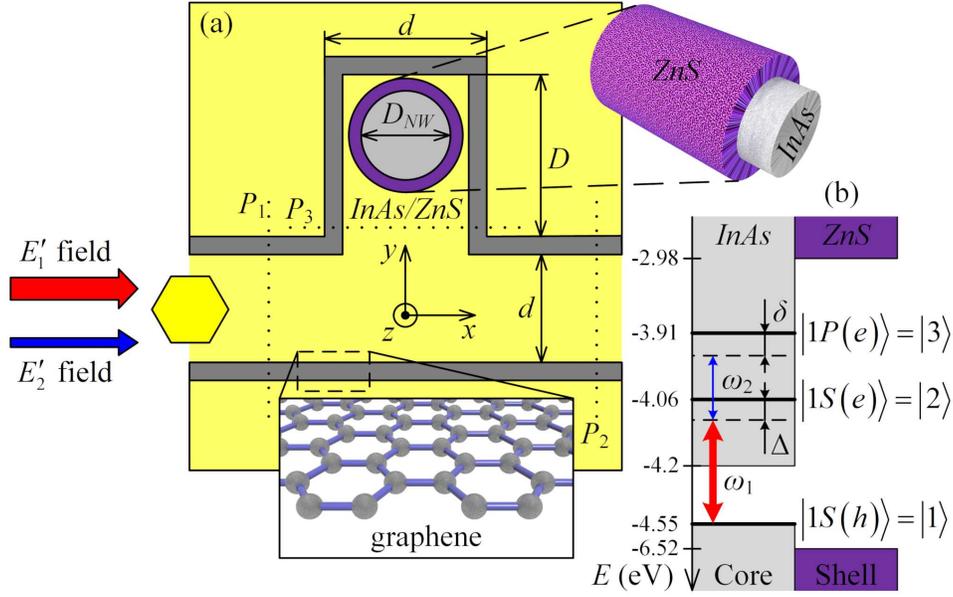}
\caption{\label{fig:7} (a) The model of graphene waveguide integrated with stub nanoresonator loaded with core-shell NW. (b) The relative position between energy gaps and band offsets of InAs-ZnS bulk semiconductors, where $E_{V1}=-4.55 \; \textrm{eV}$ for the top of the valence band and $E_{C1}=-4.2 \; \textrm{eV}$ for the bottom of the conduction band in InAs; $E_{V2}=-6.52 \; \textrm{eV}$, $E_{C2}=-2.98 \; \textrm{eV}$ the same in ZnS; the Ladder-type interaction scheme of two SPP modes with frequencies $\omega_{1}$ (pump) and $\omega_{2}$ (signal) and $9.9 \; \textrm{nm}$ core radius InAs/ZnS NW with energy levels $E_{\left|1\right\rangle}=-4.55 \; \textrm{eV}$, $E_{\left|2\right\rangle}=-4.063 \; \textrm{eV}$ and $E_{\left|3\right\rangle}=-3.908 \; \textrm{eV}$.}
\end{figure}

Now, we will investigate the graphene waveguide integrated with the stub nanoresonator (see Fig.~\ref{fig:7}) as a more complicated model for simulation. The transmittance coefficient of SPP propagating through the waveguide with stub described by equation~\cite{Lin}
\begin{equation}
\label{eq:16}
T\left(\lambda \right)=\left|t_{1}+\frac{s_{1}s_{3}}{1-r_{3}e^{i\phi \left(\lambda \right)}}e^{i\phi \left(\lambda \right)}\right|^2,
\end{equation}
where $\phi\left(\lambda\right)=\frac{2\pi \Delta S}{\lambda}$; parameters $r_{i}$, $t_{i}$, $s_{i}$ correspond to the reflection, transmission, and splitting coefficients in the \textit{i}th cross-section (\textit{i}th Ports) of the stub in Fig.~\ref{fig:7}a. Initially, we tune our waveguide to the condition of minimum transmittance, i.e., when electromagnetic mode localized by waveguide cannot pass further stub position. This setting is very simple and satisfies the requirement that the ``plasmonic path'' of mode in the stub $\Delta S=\left(2D+d\right)n^{\left(\textrm{R}\right)}_{EF+}$ (taking into account the distance between sheets in a waveguide) is a half-integral multiple of the wavelength $\frac{\left(2n+1\right)\lambda_{0}}{2}$ ($n=0,1,2\dots$), where $D$ is the height of stub nanoresonator (see Fig.~\ref{fig:7}).
Using parameters taken from Table~\ref{tab:1} for $8.04 \; \textrm{\textmu m}$ and tuning system to the minimum of \textit{0}th order, we can approximately estimate $D=23.8 \; \textrm{nm}$. The numerical simulation of the system with such parameters gives excellent evidence of our theoretical estimations. In particular, predicted characteristics (in Table~\ref{tab:1}) agree with calculated values for the strong coupling regime. The most important result, as one can see in Fig.~\ref{fig:12}a is that the SPP mode at a wavelength $\lambda_{0} =8.04 \; \textrm{\textmu m}$ is completely blocked by the stub. We can consider that the ``plasmonic transistor'' is locked under these conditions.

In this part, we consider the possibility to control the SPP propagation due to nonlinear plasmonic resonance in nanostructures~\cite{Chen}. We assume that semiconductor NW loaded into graphene stub nanoresonator interacts with two SPP modes~\cite{Koppens}, which simultaneously propagate along the pair of graphene sheets as shown in Fig.~\ref{fig:7}. The Hamiltonian of the system NW+SPPs has the following form:
\begin{subequations}
\label{eq:17}
\begin{align}
H&=H_{0}+H_{v}, \\
H_{0}&=\hbar \left(\omega_{12}\left|2\right\rangle \left\langle 2\right|+\left(\omega_{12}+\omega_{23}\right)\left|3\right\rangle \left\langle 3\right|\right), \\
H_{v}&=-\hbar \left(\widetilde{\Omega}_{1}\left|2\right\rangle \left\langle 1\right|+\widetilde{\Omega}^{*}_{1}\left|1\right\rangle \left\langle 2\right|+\widetilde{\Omega}_{2}\left|3\right\rangle \left\langle 2\right|+\widetilde{\Omega}^{*}_{2}\left|2\right\rangle \left\langle 3\right|\right),
\end{align}
\end{subequations}
where $H_{0}$ is the Hamiltonian of unexcited NW and $H_{v}$ is the Hamiltonian of interaction between NW and two SPPs with the accordance of the Ladder-type scheme in Fig.~\ref{fig:7}. Here $\left|1\right\rangle \equiv \left|1S(h)\right\rangle$ corresponds to the energy level of the hole in the valence band, $\left|2\right\rangle \equiv \left|1S(e)\right\rangle $ and $\left|3\right\rangle \equiv \left|1P(e)\right\rangle$ describe electronic levels in conduction band; $\widetilde{\Omega}_{1}$ and $\widetilde{\Omega}_{2}$ are the Rabi frequencies of pump and signal fields, respectively, $\omega_{12}$ and $\omega_{23}$ are the frequencies of interband and intraband transitions in NW, respectively.

The evolution of the presented system is described by the Liouville equation:
\begin{subequations}
\label{eq:18}
\begin{align}
\frac{\partial \widetilde{\rho}}{\partial t}&=-\frac{i}{\hbar}\left[H,\widetilde{\rho}\right]-\widehat{\Gamma}, \\
\nonumber
\widetilde{\rho}&=\widetilde{\rho}_{11}\left|1\right\rangle \left\langle 1\right|+\widetilde{\rho}_{22}\left|2\right\rangle \left\langle 2\right|+\widetilde{\rho}_{33}\left|3\right\rangle \left\langle 3\right|+\widetilde{\rho}_{12}\left|1\right\rangle \left\langle 2\right|+\widetilde{\rho}_{21}\left|2\right\rangle \left\langle 1\right| \\
&\mspace{20mu}+\widetilde{\rho}_{23}\left|2\right\rangle \left\langle 3\right|+\widetilde{\rho}_{32}\left|3\right\rangle \left\langle 2\right|+\widetilde{\rho}_{13}\left|1\right\rangle \left\langle 3\right|+\widetilde{\rho}_{31}\left|3\right\rangle \left\langle 1\right|, \\
\nonumber
\widehat{\Gamma}&=\gamma_{21}\left(\left|2\right\rangle \left\langle 2\right|\widetilde{\rho}-2\left|1\right\rangle \left\langle 2\right|\widetilde{\rho}\left|2\right\rangle \left\langle 1\right|+\widetilde{\rho}\left|2\right\rangle \left\langle 2\right|\right)+\gamma_{32}\left(\left|3\right\rangle \left\langle 3\right|\widetilde{\rho}-2\left|2\right\rangle \left\langle 3\right|\widetilde{\rho}\left|3\right\rangle \left\langle 2\right|+\widetilde{\rho}\left|3\right\rangle \left\langle 3\right|\right) \\
&\mspace{20mu}+\gamma_{31}\left(\left|3\right\rangle \left\langle 3\right|\widetilde{\rho}-2\left|1\right\rangle \left\langle 3\right|\widetilde{\rho}\left|3\right\rangle \left\langle 1\right|+\widetilde{\rho}\left|3\right\rangle \left\langle 3\right|\right),
\end{align}
\end{subequations}
where $\widetilde{\rho}$ is the density matrix for energy levels in NW, $\widehat{\Gamma}$ is the Lindblad superoperator describing the processes of spontaneous relaxation in the system, ${\gamma }_{ij}$ are the spontaneous relaxation rates for corresponding transitions, $i,j=1,2,3$ and $i\neq j$.

Using (\ref{eq:17})--(\ref{eq:18}) it is possible to obtain the system of equations for the evolution of density matrix elements:
\begin{subequations}
\label{eq:19}
\begin{align}
\dot{\widetilde{\rho}}_{11}&=i\widetilde{\Omega}^{*}_{1}\widetilde{\rho}_{21}-i\widetilde{\Omega}_{1}\widetilde{\rho}_{12}
+2\gamma_{21}\widetilde{\rho}_{22}+2\gamma_{31}\widetilde{\rho}_{33}, \\
\dot{\widetilde{\rho}}_{22}&=i\widetilde{\Omega}_1\widetilde{\rho}_{12}-i\widetilde{\Omega}^{*}_{1}\widetilde{\rho}_{21}
+i\widetilde{\Omega}^{*}_{2}\widetilde{\rho}_{32}-i\widetilde{\Omega}_{2}\widetilde{\rho}_{23}-2\gamma_{21}\widetilde{\rho}_{22}
+2\gamma_{32}\widetilde{\rho}_{33}, \\
\dot{\widetilde{\rho}}_{33}&=i\widetilde{\Omega}_{2}\widetilde{\rho}_{23}-i\widetilde{\Omega}^{*}_{2}\widetilde{\rho}_{32}
-2\gamma_{32}\widetilde{\rho}_{33}-2\gamma_{31}\widetilde{\rho}_{33}, \\
\dot{\widetilde{\rho}}_{12}&=i\widetilde{\Omega}^{*}_{1}\widetilde{\rho}_{22}+i\omega_{12}\widetilde{\rho}_{12}
-i\widetilde{\Omega}^{*}_{1}\widetilde{\rho}_{11}-i\widetilde{\Omega}_{2}\widetilde{\rho}_{13}-\gamma_{21}\widetilde{\rho}_{12}, \\
\dot{\widetilde{\rho}}_{21}&=-i\widetilde{\Omega}_{1}\widetilde{\rho}_{22}-i\omega_{12}\widetilde{\rho}_{21}
+i\widetilde{\Omega}_{1}\widetilde{\rho}_{11}+i\widetilde{\Omega}^{*}_{2}\widetilde{\rho}_{31}-\gamma_{21}\widetilde{\rho}_{21}, \\
\dot{\widetilde{\rho}}_{13}&=i\widetilde{\Omega}^{*}_{1}\widetilde{\rho}_{23}+i\left(\omega_{12}
+\omega_{23}\right)\widetilde{\rho}_{13}-i\widetilde{\Omega}^{*}_{2}\widetilde{\rho}_{12}
-\gamma_{31}\widetilde{\rho}_{13}-\gamma_{32}\widetilde{\rho}_{13}, \\
\dot{\widetilde{\rho}}_{31}&=-i\widetilde{\Omega}_{1}\widetilde{\rho}_{32}-i\left(\omega_{12}
+\omega_{23}\right)\widetilde{\rho}_{31}+i\widetilde{\Omega}_{2}\widetilde{\rho}_{21}
-\gamma_{31}\widetilde{\rho}_{31}-\gamma_{32}\widetilde{\rho}_{31}, \\
\dot{\widetilde{\rho}}_{23}&=i\omega_{23}\widetilde{\rho}_{23}+i\widetilde{\Omega}_{1}\widetilde{\rho}_{13}
+i\widetilde{\Omega}^{*}_{2}\widetilde{\rho}_{33}-i\widetilde{\Omega}^{*}_{2}\widetilde{\rho}_{22}
-\widetilde{\rho}_{23}\left(\gamma_{21}+\gamma_{32}+\gamma_{31}\right), \\
\dot{\widetilde{\rho}}_{32}&=-i\omega_{23}\widetilde{\rho}_{32}-i\widetilde{\Omega}^{*}_{1}\widetilde{\rho}_{31}
-i\widetilde{\Omega}_{2}\widetilde{\rho}_{33}+i\widetilde{\Omega}_{2}\widetilde{\rho}_{22}
-\widetilde{\rho}_{32}\left(\gamma_{21}+\gamma_{32}+\gamma_{31}\right).
\end{align}
\end{subequations}
We use the approximation of slowly varying amplitudes for passing to the new variables:
\begin{gather}
\nonumber
\widetilde{\rho}_{12}=\rho_{12}e^{i\omega_{1}t}, \; \widetilde{\rho}_{23}=\rho_{23}e^{i\omega_{2}t}, \; \widetilde{\rho}_{13}=\rho_{13}e^{i\left(\omega_{1}+\omega_{2}\right)t}, \\
\nonumber
\widetilde{\rho}_{11} \equiv \rho_{11}, \; \widetilde{\rho}_{22} \equiv \rho_{22}, \; \widetilde{\rho}_{33} \equiv \rho_{33}, \;
\widetilde{\Omega}_{1}=\Omega_{1}e^{i\omega_{1}t}, \; \widetilde{\Omega}_{2}=\Omega_{2}e^{i\omega_{2}t},
\end{gather}
where $\omega_{1\left(2\right)}$ is the frequency of the pump (signal) field. The system of Equations (\ref{eq:19}) transforms into a new form:
\begin{subequations}
\label{eq:20}
\begin{align}
\dot{\rho}_{11}&=i\Omega^{*}_{1}\rho_{21}-i\Omega_{1}\rho_{12}+2\gamma_{21}\rho_{22}+2\gamma_{31}\rho_{33}, \\
\dot{\rho}_{22}&=i\Omega_{1}\rho_{12}-i\Omega^{*}_{1}\rho_{21}+i\Omega^{*}_{2}\rho_{32}-i\Omega_{2}\rho_{23}
-2\gamma_{21}\rho_{22}+2\gamma_{32}\rho_{33}, \\
\dot{\rho}_{33}&=i\Omega_{2}\rho_{23}-i\Omega^{*}_{2}\rho_{32}-2\gamma_{32}\rho_{33}-2\gamma_{31}\rho_{33}, \\
\dot{\rho}_{12}&=i\Omega^{*}_{1}\rho_{22}+i\Delta \rho_{12}-i\Omega^{*}_{1}\rho_{11}-i\Omega_{2}\rho_{13}-\gamma_{21}\rho_{12}, \\
\dot{\rho}_{21}&=-i\Omega_{1}\rho_{22}-i\Delta \rho_{21}+i\Omega_{1}\rho_{11}+i\Omega^{*}_{2}\rho_{31}-\gamma_{21}\rho_{21}, \\
\dot{\rho}_{13}&=i\Omega^{*}_{1}\rho_{23}+i\delta \rho_{13}-i\Omega^{*}_{2}\rho_{12}-\gamma_{31}\rho_{13}-\gamma_{32}\rho_{13}, \\
\dot{\rho}_{31}&=-i\Omega_{1}\rho_{32}-i\delta \rho_{31}+i\Omega_{2}\rho_{21}-\gamma_{31}\rho_{31}-\gamma_{32}\rho_{31}, \\
\dot{\rho}_{23}&=i\left(\delta -\Delta\right)\rho_{23}+i\Omega_{1}\rho_{13}+i\Omega^{*}_{2}\rho_{33}-i\Omega^{*}_{2}\rho_{22}
-\rho_{23}\left(\gamma_{21}+\gamma_{32}+\gamma_{31}\right), \\
\dot{\rho}_{32}&=-i\left(\delta -\Delta \right)\rho_{32}-i\Omega^{*}_{1}\rho_{31}-i\Omega_{2}\rho_{33}+i\Omega_{2}\rho_{22}
-\rho_{32}\left(\gamma_{21}+\gamma_{32}+\gamma_{31}\right).
\end{align}
\end{subequations}
where $\Delta=\omega_{12}-\omega_{1}$, $\delta =\omega_{12}+\omega_{23}-\omega_{1}-\omega_{2}$. Defining new variables, we represent the system (\ref{eq:20}) in the following form:
\begin{subequations}
\label{eq:21}
\begin{align}
\dot{n}_{21}&=2i\Omega_{1}\rho_{12}-2i\Omega^{*}_{1}\rho_{21}+i\Omega^{*}_{2}\rho_{32}-i\Omega_{2}\rho_{23}
-4\gamma_{21}\rho_{22}+2\left(\gamma_{32}-\gamma_{31}\right)\rho_{33}, \\
\dot{n}_{32}&=2i\Omega_{2}\rho_{23}-2i\Omega^{*}_{2}\rho_{32}-i\Omega_{1}\rho_{12}+i\Omega^{*}_{1}\rho_{21}
-4\gamma_{32}\rho_{33}-2\gamma_{31}\rho_{33}+2\gamma_{21}\rho_{22}, \\
\dot{\rho}_{21}&=-i\Omega_{1}n_{21}-i\Delta \rho_{21}+i\Omega^{*}_{2}\rho_{31}-\gamma_{21}\rho_{21}, \\
\dot{\rho}_{32}&=-i\Omega_{2}n_{32}-i\left(\delta -\Delta\right)\rho_{32}-i\Omega^{*}_{1}\rho_{31}-\left(\gamma_{21}+\gamma_{32}+\gamma_{31}\right)\rho_{32}, \\
\dot{\rho}_{31}&=-i\Omega_{1}\rho_{32}-i\delta \rho_{31}+i\Omega_{2}\rho_{21}-\left(\gamma_{31}+\gamma_{32}\right)\rho_{31},
\end{align}
\end{subequations}
where $n_{21}=\rho_{22}-\rho_{11}$, $n_{32}=\rho_{33}-\rho_{22}$.

In case when the system reaches the stationary regime (i.e. $\dot{n}_{21}=\dot{n}_{32}=\dot{\rho}_{21}=\dot{\rho}_{31}=\dot{\rho}_{32}=0$) the polarization and population imbalances have the steady-state values. In particular, we express $\rho_{31}$ from Equation (\ref{eq:21}e)
\begin{equation}
\label{eq:22}
\overline{\rho}_{31}=\frac{i\Omega_{2}\overline{\rho}_{21}-i\Omega_{1}\overline{\rho}_{32}}{i\delta +\gamma_{31}+\gamma_{32}},
\end{equation}
where $\overline{\rho}_{21}$, $\overline{\rho}_{32}$ and $\overline{\rho}_{31}$ are the stationary values of polarizations for corresponding transitions. We substitute $\overline{\rho}_{31}$ into (\ref{eq:21}c) and (\ref{eq:21}d) and obtain
\begin{subequations}
\label{eq:23}
\begin{align}
0&=-i\Omega_{1}n_{21}-\rho_{21}\left(i\Delta+\gamma_{21}+\frac{\left|\Omega_{2}\right|^{2}}{i\delta +\gamma_{31}+\gamma_{32}}\right)+\frac{\Omega_{1}\Omega^{*}_{2}\rho_{32}}{i\delta +\gamma_{31}+\gamma_{32}}, \\
0&=-i\Omega_{2}n_{32}+\frac{\Omega^{*}_{1}\Omega_{2}\rho_{21}}{i\delta +\gamma_{31}+\gamma_{32}}-\rho_{32}\left(\gamma_{21}+\gamma_{32}+\gamma_{31}+i\left(\delta -\Delta\right)+\frac{\left|\Omega_{1}\right|^{2}}{i\delta +\gamma_{31}+\gamma_{32}}\right).
\end{align}
\end{subequations}
Solving the system (\ref{eq:23}) we find stationary solutions for $\overline{\rho}_{21}$ and $\overline{\rho}_{32}$ in the following form
\begin{subequations}
\label{eq:24}
\begin{align}
\overline{\rho}_{21}&=-\frac{i\Omega_{1}\left(\left|\Omega_{1}\right|^{2}\overline{n}_{21}
+\left|\Omega_{2}\right|^{2}\overline{n}_{32}+D_{2}\overline{n}_{21}\Gamma_{32}\right)}
{\left|\Omega_{1}\right|^{2}D_{1}+D_{1}D_{2}\Gamma_{32}+\left|\Omega_{2}\right|^{2}\Gamma_{32}}, \\
\overline{\rho}_{32}&=-\frac{i\Omega_{2}\left(\left|\Omega_{1}\right|^{2}\overline{n}_{21}+D_{1}D_{2}\overline{n}_{32}
+\left|\Omega_{2}\right|^{2}\overline{n}_{32}\right)}{\left|\Omega_{1}\right|^{2}D_{1}+D_{1}D_{2}\Gamma_{32}
+\left|\Omega_{2}\right|^{2}\Gamma_{32}},
\end{align}
\end{subequations}
where $D_{1}=i\Delta+\gamma_{21}$, $D_{2}=i\delta +\gamma_{31}+\gamma_{32}$, $\Gamma_{32}=i\left(\delta -\Delta\right)+\gamma_{21}+\gamma_{31}+\gamma_{32}$, $\overline{n}_{21}=\overline{\rho}_{22}-\overline{\rho}_{11}$, $\overline{n}_{32}=\overline{\rho}_{33}-\overline{\rho}_{22}$; $\overline{\rho}_{11}$, $\overline{\rho}_{22}$ and $\overline{\rho}_{33}$ are the stationary values of populations for the corresponding energy levels. Substituting Eqs. (\ref{eq:24}) into the system (\ref{eq:20}) and solving it, we can find the stationary solutions for the populations of energy levels as follows:
\begin{subequations}
\label{eq:25}
\begin{align}
\overline{\rho}_{11}&=1-\overline{\rho}_{22}-\overline{\rho}_{33}, \\
\overline{\rho}_{33}&=\frac{\left|\Omega_{2}\right|^{2}\left|\Omega_{1}\right|^{2}}{A}\left(\left|\Omega_{2}\right|^{2}\Gamma^{2}_{1}
+\left(\left(\delta -\Delta\right)^{2}+\Gamma^{2}_{1}\right)\Gamma_{2}\gamma_{21}+\Gamma_{1}\gamma_{21}\left|\Omega_{1}\right|^{2}\right),\\
\nonumber
\overline{\rho}_{22}&=\frac{\left|\Omega_{1}\right|^{2}}{A}\Bigl(\left|\Omega_{2}\right|^{4}\Gamma_{1}\gamma_{32}
+\left|\Omega_{2}\right|^{2}\Bigl(\delta^{2}\left(\Gamma^{2}_{2}+\left(\Gamma_{2}+\Gamma_{3}\right)\gamma_{21}\right)-2\delta \Delta\Gamma_{2}\gamma_{32} \\
&\mspace{20mu}+\Gamma_{2}\left(\Gamma^{2}_{1}\Gamma_{2}+\Delta^{2}\gamma_{32}+\Gamma_{1}\gamma_{21}\gamma_{32}\right)
+\left(\Gamma_{1}\Gamma_{2}+\gamma_{21}\gamma_{32}\right)\left|\Omega_{1}\right|^{2}\Bigr)+B\Bigr),
\end{align}
\end{subequations}
where
\begin{align}
\nonumber
A&=\left|\Omega_{2}\right|^{6}\Gamma_{1}\Gamma_{3}+B\left(\Delta^{2}+\gamma^{2}_{21}+2\left|\Omega_{1}\right|^{2}\right)
+\left|\Omega_{2}\right|^{4}\Bigl(\left(\delta^{2}+\Delta^{2}+\Gamma_{1}\left(\Gamma_{1}
+2\Gamma_{3}\right)\right)\Gamma_{2}\gamma_{21} \\
\nonumber
&\mspace{20mu}-2\delta \Delta\left(\Gamma_{1}\Gamma_{3}+\Gamma_{2}\gamma_{21}\right)
+\left(\gamma^{2}_{21}+\left(\Gamma_{1}+\Gamma_{2}\right)\left(\Gamma_{1}-\gamma_{31}\right)\right)
\left|\Omega_{1}\right|^{2}\Bigr) \\
\nonumber
&\mspace{20mu}+\left|\Omega_{2}\right|^{2}\Bigl(\delta^{2}\left(\left(2\Gamma^{2}_{2}
+\Gamma_{1}\Gamma_{3}\right)\gamma^{2}_{21} +\Delta^{2}\left(\Gamma_{1}\Gamma_{3}+4\Gamma_{2}\gamma_{21}\right)
+2\left(\Gamma^{2}_{1}+\gamma_{21}\gamma_{31}\right)\left|\Omega_{1}\right|^{2}\right) \\
\nonumber
&\mspace{20mu}+2\delta \Delta\Gamma_{2}\left(-\gamma_{21}\left(\delta^{2}+\Delta^{2}+\Gamma^{2}_{1}+2\Gamma_{2}\gamma_{21}\right)
+\left(\gamma_{31}-\gamma_{32}\right)\left|\Omega_{1}\right|^{2}\right) \\
\nonumber
&\mspace{20mu}+\Delta^{2}\Gamma_{2}\left(\Gamma_{1}\Gamma_{2}\Gamma_{3}+2\Gamma_{2}\gamma^{2}_{21}
+\gamma_{32}\left|\Omega_{1}\right|^{2}\right) +\left(\Gamma_{1}\Gamma_{2}+\left|\Omega_{1}\right|^{2}\right) \\
\nonumber
&\mspace{20mu}\times\left(\Gamma_{2}
\left(2\Gamma_{1}+\Gamma_{3}\right)\gamma^{2}_{21}+\left(\Gamma^{2}_{1}+\Gamma^{2}_{2}+2\gamma_{21}\gamma_{32}\right)
\left|\Omega_{1}\right|^{2}\right)\Bigr), \\
\nonumber
B&=\Gamma_{2}\gamma_{21}\left(\left(\left(\delta -\Delta\right)^{2}+\Gamma^{2}_{1}\right)\left(\delta^{2}+\Gamma^{2}_{2}\right)+2\left(\delta \left(\Delta-\delta\right)
+\Gamma_{1}\Gamma_{2}\right)\left|\Omega_{1}\right|^{2}+\left|\Omega_{1}\right|^{4}\right), \\
\nonumber
\Gamma_{1}&=\gamma_{21}+\gamma_{31}+\gamma_{32}, \; \Gamma_{2}=\gamma_{31}+\gamma_{32}, \; \Gamma_{3}=\gamma_{21}+\gamma_{31}.
\end{align}

We need to carry out the correctness and stability analysis of our stationary solutions. Initially, we substitute the fixed values of material parameters into (\ref{eq:24})--(\ref{eq:25}) and change (optimize) the intensities of signal and pump SPPs (and field detunings) in order to achieve the stationary regime of the system with physically realizable parameter values of populations and polarizations $n_{21}$, $n_{32}$, $\rho_{21}$, $\rho_{31}$, $\rho_{32}$. Next, we numerically simulate system (\ref{eq:20}) with initial values of matrix elements $\rho_{11}=1$, $\rho_{22}=\rho_{33}=\rho_{21}=\rho_{31}=\rho_{32}=0$ and find that after evolution, all matrix elements reach the stationary values (\ref{eq:22}), (\ref{eq:24})--(\ref{eq:25}). Note that expressions (\ref{eq:24}) can be used independently of the solutions (\ref{eq:25}) if we initially know the values of level populations satisfying to the stationary regime in the scheme. Thus, we realized the stress-test of our numerical solutions using the deviation of the initial values of the density matrix elements from stationary values and proved the stability of our stationary solutions.

Besides, we are interested in the contribution of various nonlinear processes to the formation of stationary propagation regimes of a signal SPP. For this purpose, we substitute (\ref{eq:22}) into (\ref{eq:21}d) and obtain the following equation for the evolution of the density matrix element corresponding to polarization on signal transition:
\begin{equation}
\label{eq:26}
\dot{\rho}_{32}=-i\Omega_{2}n_{32}+\frac{\Omega^{*}_{1}\Omega_{2}\rho_{21}}{i\delta +\gamma_{31}+\gamma_{32}}-\frac{\left|\Omega_{1}\right|^{2}\rho_{32}}{i\delta +\gamma_{31}+\gamma_{32}}-\rho_{32}\left(\gamma_{21}+\gamma_{31}+\gamma_{32}+i\left(\delta -\Delta\right)\right).
\end{equation}
This representation is a power series expansion in the Rabi frequencies of the signal and pump SPPs. The expression (\ref{eq:26}) can be represented in the form that is convenient for the further analysis of various terms contribution into system dynamics:
\begin{equation}
\label{eq:27}
\dot{\rho}_{32}=\sum^{4}_{i=1}{X_{i}},
\end{equation}
where $X_{1}=-i\Omega_{2}n_{32}$ corresponds to the induced single-quantum transitions in the system, $X_{2}=\frac{\Omega^{*}_{1}\Omega_{2}\rho_{21}}{i\delta +\gamma_{31}+\gamma_{32}}$ corresponds to the nonlinear scattering, $X_{3}=-\frac{\left|\Omega_{1}\right|^{2}\rho_{32}}{i\delta +\gamma_{31}+\gamma_{32}}$ corresponds to the cross-interaction between SPPs and $X_{4}=-\rho_{32}\left(\gamma_{21}+\gamma_{31}+\gamma_{32}+i\left(\delta -\Delta\right)\right)$ corresponds to the linear effects associated with the dispersion and spontaneous decay of the excited states. The estimation of the contribution of various effects into graphene device functioning in the stationary regime is shown in Table~\ref{tab:2}.
\begin{table}[H]
\caption{The contribution of various effects into the formation of the stationary regime for signal SPP.}
\label{tab:2}
\centering
\footnotesize
\begin{tabular}{*{7}{c}}
\toprule
$\rho_{21}$ & $\rho_{32}$ & $\rho_{31}$ & $X_{1}$ & $X_{2}$ & $X_{3}$ & $X_{4}$ \\
\midrule
$\overline{\rho}_{21}$ & $\overline{\rho}_{32}$ & $\overline{\rho}_{31}$ & $1.56\cdot {10}^{12}i$ & $1.88\cdot {10}^{10}-6.86\cdot {10}^{10}i$ & $-4.87\cdot {10}^{11}+1.51\cdot {10}^{12}i$ & $4.68\cdot {10}^{11}-2.00\cdot {10}^{12}i$ \\
\bottomrule
\end{tabular}
\end{table}

\section{Tuning the NW size to satisfy the resonance conditions for intraband and interband transitions induced by signal and pump SPPs}
We start with tuning intraband transition $1S\left(e\right) \rightarrow 1P\left(e\right)$ in core-shell NW to the wavelength $\lambda_{2}$ for signal SPP supported by a pair of graphene sheets under the condition that interband transition $1S\left(h\right) \rightarrow 1S\left(e\right)$ is tuned to the wavelength $\lambda_{1}$ for pump SPP supported by a graphene waveguide too. The suitable active center for this purpose is the InAs/ZnS core-shell NW~\cite{Chen2,Harrison,Bouarissa,Cao1}. The parameters of such NW taken from the literature are summarized in Table~\ref{tab:3}. The information about the position of the energy levels is presented in Table~\ref{tab:4}.

We assume that both NWs (source and NW inside of the stub) cannot support propagating guided modes, but the near-field interaction regime corresponds to the generation of leaky modes~\cite{Grzela}. Since the $z$-guided modes are not supported by NW, to calculate the corresponding wavelengths of intraband and interband transitions we use the following equations~\cite{Harrison}
\begin{subequations}
\label{eq:28}
\begin{align}
\omega_{12}&=\frac{eE_{g}}{\hbar}+\frac{2\hbar \kappa^{2}_{1,0}}{D^{2}_{NW}}\left(\frac{1}{m_{c}}+\frac{1}{m_{h}}\right),\\
\omega_{23}&=\frac{2\hbar}{D^{2}_{NW}m_{c}}\left(\kappa^{2}_{1,1}-\kappa^{2}_{1,0}\right),
\end{align}
\end{subequations}
where $E_{g}$ is the band gap of the semiconductor, $m_{c}$ and $m_{h}$ are the effective masses of electron and hole, respectively, $\kappa_{1,1}=4.493$ and $\kappa_{1,0}=\pi$ are the roots of the Bessel function, $a_{NW}=D_{NW}/2$ is the radius of the NW core.

\begin{table}[H]
\caption{The parameters of core-shell NW and transitions in it.}
\label{tab:3}
\centering
\footnotesize
\begin{tabular}{*{10}{c}}
\toprule
\multirow{2}{*}{\makecell{Semiconductor \\ material}} & \multirow{2}{*}{$\varepsilon$} & \multirow{2}{*}{$m_{c}$, $m_{0}$} & \multirow{2}{*}{$m_{h}$, $m_{0}$} & \multirow{2}{*}{$E_{g}, \; \textrm{eV}$} & \multirow{2}{*}{$a_{NW}, \; \textrm{nm}$} & \multicolumn{2}{c}{$1S\left(e\right) \rightarrow 1P\left(e\right)$} & \multicolumn{2}{c}{$1S\left(h\right) \rightarrow 1S\left(e\right)$} \\
& & & & & & $\lambda_{2}, \; \textrm{\textmu m}$ & $\mu_{32}, \; \textrm{C} \cdot \textrm{m}$ & $\lambda_{1}, \; \textrm{\textmu m}$ & $\mu_{12}, \; \textrm{C} \cdot \textrm{m}$ \\
\midrule
core, InAs & $12.3$ & $0.026$ & $0.41$ & $0.35$ & $9.9$ & $8.04$ & $5.91\cdot 10^{-28}$ & $2.56$ & $14.9\cdot 10^{-29}$ \\
shell, ZnS & $8.3$ & $0.27$ & $0.58$ & $3.54$ & $10$ & \multicolumn{4}{c}{} \\
\bottomrule
\end{tabular}
\end{table}

The dipole moment of the interband transition is calculated in accordance with the formula~\cite{mork}
\begin{equation}
\label{eq:29}
\mu^{2}_{12}=\frac{e^{2}}{6m_{0}\omega^{2}_{1}}\left(\frac{m_{0}}{m_{c}}-1\right)\frac{E_{g}e\left(E_{g}
+\Delta_{s}\right)}{E_{g}+2\Delta_{s}/3},
\end{equation}
where $\Delta_{s}$ is the spin-orbit splitting for the material of NW core ($\Delta_{s}=0.43 \; \textrm{eV}$), $m_{0}$ is the free-electron mass. More complicated formulas are required to calculate the dipole moment of the intraband transition, but they can be approximated by the expression $\mu_{32}=0.433ea_{NW}\Lambda$, where $\Lambda=3\varepsilon_{ZnS}/\left(2\varepsilon_{ZnS}+\varepsilon_{InAs}\right)$. Using NW radius $9.9 \; \textrm{nm}$, we get the wavelength $\lambda_{2}=8.04 \; \textrm{\textmu m} $ for signal SPP and $\lambda_{1}=2.56 \; \textrm{\textmu m}$ for pump SPP that are simultaneously supported by graphene waveguide with $\mu_{c}=0.6 \; \textrm{eV}$, $\tau =0.9 \; \textrm{ps}$. The other working interaction parameters were obtained and summarized in Table~\ref{tab:3}.

\begin{table}[H]
\caption{The bands and energy levels positions in InAs/ZnS core-shell NW.}
\label{tab:4}
\centering
\footnotesize
\begin{tabular}{*{6}{c}}
\toprule
\makecell{Semiconductor \\ material} & \makecell{Top of the valence \\ band $E_{V}, \; \textrm{eV}$} & \makecell{Bottom of the conduction \\ band $E_{C}, \; \textrm{eV}$} & \makecell{Energy level \\ $E_{\left|1\right\rangle}, \; \textrm{eV}$} & \makecell{Energy level \\ $E_{\left|2\right\rangle}, \; \textrm{eV}$} & \makecell{Energy level \\ $E_{\left|3\right\rangle}, \; \textrm{eV}$} \\
\midrule
core, InAs & $-4.55$ & $-4.2$ & $-4.55$ & $-4.063$ & $-3.908$ \\
shell, ZnS & $-6.52$ & $-2.98$ & \multicolumn{3}{c}{} \\
\bottomrule
\end{tabular}
\end{table}

\section{Local density of states and modification of relaxation rate and coupling constant of NW at a nanoscale distance to graphene}
\begin{figure}[t]
\centering
\includegraphics[width=0.5\columnwidth]{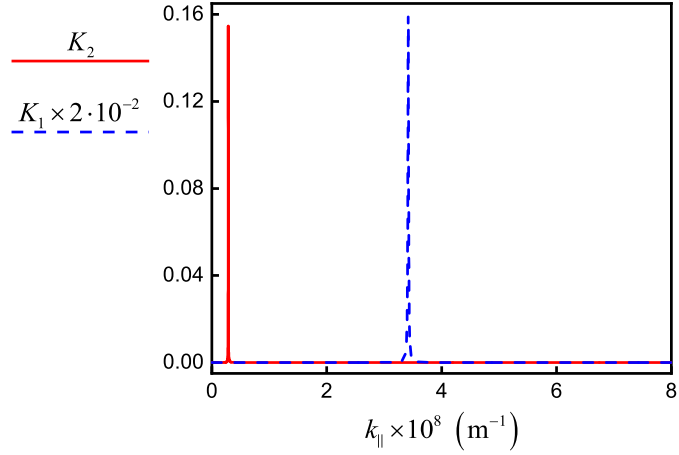}
\caption{\label{fig:8} The dependence of integrands $K_{1}$ for wavelength $\lambda_{1}=2.56 \; \textrm{\textmu m}$ (dashed blue line) and $K_{2}$ for wavelength $\lambda_{2}=8.04 \; \textrm{\textmu m}$ (solid red line) as a function of the in-plane wave vector $k_{\parallel}$ for emitting InAs/ZnS NW that is placed at $10 \; \textrm{nm}$ from the graphene (parameters in Table~\ref{tab:1}).}
\end{figure}

The emitter relaxation rate can change due to a modification in the local density of plasmonic states of the self-consistent field, for example, when the emitter is placed in a resonator. In the beginning, we consider the simplest case when the emitter is located near the flat conductive surface~\cite{Koppens,Chance,Weber,Larkin}. We introduce a set of parameters $\kappa =\frac{\Gamma}{\Gamma_{0}}$, $\kappa_{SPP}=\frac{\Gamma_{SPP}}{\Gamma_{0}}$, $\kappa_{SP}=\frac{\Gamma_{SP}}{\Gamma_{0}}$, $\kappa_{L}=\frac{\Gamma_{L}}{\Gamma_{0}}$, which describe the change in relaxation rate of the emitter, where $\Gamma=\Gamma_{0}+\Gamma_{0}\int^{\infty}_{0}{K\left(k_{\parallel}\right)\textrm{d}k_{\parallel}}$ is the total rate of relaxation, $\Gamma_{SP}=\Gamma_{0}\int^{\infty}_{k_{1}}{K\left(k_{\parallel}\right)\textrm{d}k_{\parallel}}$ is the SP-mediated rate of evanescent waves generation, $\Gamma_{SPP}=\Gamma_{0}\int^{k_{SPP}+\Delta k}_{k_{SPP}-\Delta k}{K\left(k_{\parallel}\right)\textrm{d}k_{\parallel}}$ is the relaxation rate of propagated SPPs and $\Gamma_{L}=\Gamma_{0}+\Gamma_{0}\int^{k_{1}}_{0}{K\left(k_{\parallel}\right)\textrm{d}k_{\parallel}}$ is the radiative relaxation rate, $\Gamma_{0}\equiv \gamma^{\left(0\right)}_{ij}$ is the relaxation rate of an isolated emitter for the corresponding transition. Here $K\left(k_{\parallel}\right)=\frac{3}{4}\textrm{Re}\left\{\left(\frac{\left|\mu_{\parallel}\right|^{2}}{\left|\mu\right|^2}r^{s}
-\frac{\left|\mu_{\parallel}\right|^{2}}{\left|\mu\right|^{2}}r^{p}\left(1-\frac{k^{2}_{\parallel}}{k^{2}_{1}}\right)
+2\frac{\left|\mu_{\perp}\right|^{2}}{\left|\mu\right|^{2}}r^{p}\frac{k^{2}_{\parallel}}{k^{2}_{1}}\right)
\frac{k_{\parallel}}{k_{z1}k_{1}}e^{2ik_{z1}z_{0}}\right\}$~\cite{Koppens,Novotny} depends on both the NW-graphene distance $z_{0}$ and on the NW radius by (\ref{eq:28}), where $k_{i}=\left|k_{0}\textrm{Re}\left(n_{i}\left(\omega\right)\right)\right|$ is the absolute value of wave vector in \textit{i}th medium with refractive index $n_{i}\left(\omega\right)$, $k_{\parallel}$ is the in-plane wave vector, $\mu_{\parallel}$ and $\mu_{\perp}$ are the components of the transition dipole parallel and perpendicular to the graphene plane, $r^{p\left(s\right)}=\frac{r^{p\left(s\right)}_{1,2}+r^{p\left(s\right)}_{2,3}e^{2ik_{z2}d_{gr}}}
{1+r^{p\left(s\right)}_{1,2}r^{p\left(s\right)}_{2,3}e^{2ik_{z2}d_{gr}}}$ are the generalized Fresnel reflection coefficients for \textit{p}- and \textit{s}-polarized plane waves of a single layer of thickness $d_{gr}$ (we take graphene thickness $d_{gr}=0.33 \; \textrm{nm}$), $r^{p}_{i,j}=\frac{\varepsilon_{j}k_{zi}-\varepsilon_{i}k_{zj}}{\varepsilon_{j}k_{zi}+\varepsilon_{i}k_{zj}}$ and $r^{s}_{i,j}=\frac{\mu_{j}k_{zi}-\mu_{i}k_{zj}}{\mu_{j}k_{zi}+\mu_{i}k_{zj}}$ are the Fresnel reflection coefficients for \textit{p}- and \textit{s}-polarized plane waves, respectively, for a single interface $\left(i,j\right)$ with the medium of light incidence denoted by $i$, $k_{zi}=\sqrt{k^{2}_{i}-k^{2}_{\parallel}}$; $\varepsilon_{i}$ and $\mu_{i}$  are the permittivity and magnetic permeability, respectively, $i,j=1,2,3$. Index $1$ corresponds to the dielectric layer with NW, $2$ corresponds to the graphene layer, and $3$ corresponds to the dielectric layer without NW. In our calculations, we use $\varepsilon_{1}=\varepsilon_{3}=\varepsilon_{d}$, $\varepsilon_{2}=\varepsilon_{gr}$ and $\mu_{1}=\mu_{2}=\mu_{3}=1$, $\mu_{\parallel}=\mu_{\perp}=\mu_{12\left(32\right)}$, where in formula (\ref{eq:6}) for $\varepsilon_{gr}$ we change the effective thickness of graphene $\Delta_{g}$ on its real thickness $d_{gr}$. Figure~\ref{fig:8} shows the dependence of the integrand $K\left(k_{\parallel}\right)$ as a function of the scattered field wave vector. Plasmon peaks in Fig.~\ref{fig:8} are seen as sharp peaks near the wave vectors $k_{SPP}\left(\lambda_{1}\right)$ and $k_{SPP}\left(\lambda_{2}\right)$ of SPPs, which correspond to the wavelengths $\lambda_{1}$ for the pump and $\lambda_{2}$ for signal incident fields. With the selected parameters, the $K$ function does not have other peaks, so we choose $\Delta k=k_{SPP}-k_{1}$.

\begin{figure}[t]
\centering
\includegraphics[width=0.5\columnwidth]{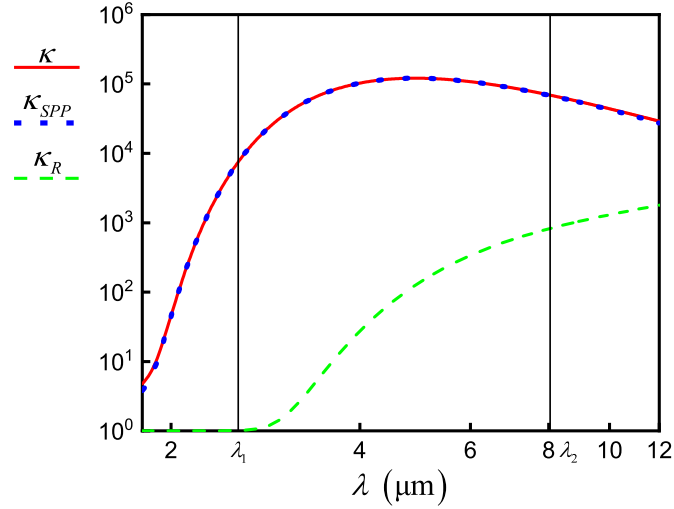}
\caption{\label{fig:9} The dependence of relaxation parameters $\kappa$ (solid red line), $\kappa_{SPP}$ (dotted blue line) and $\kappa_{R}$ (dashed green line) as a function of incident field wavelength.}
\end{figure}

In full representation of the problem like (\ref{eq:18}), we can separate the coherent processes of SPP-NW interaction in Hamiltonian and all other relaxation processes in Lindblad superoperator. The second corresponds to the relaxation parameter $\kappa_{R}=\frac{\Gamma -\Gamma_{SPP}}{\Gamma_{0}}$ that is obtained from the law of energy conservation $\kappa_{R}+\kappa_{SPP}=\kappa_{L}+\kappa_{SP}=\kappa$. Figure~\ref{fig:9} demonstrates the giant enhancement of relaxation rate for distance $z_{0}=10 \; \textrm{nm}$ between the center of NW and graphene in the selected wavelength range. We note that the dominant part of the excited NW energy is distributed to SPP generation. The contribution of other processes to the relaxation acceleration is presented in Fig.~\ref{fig:9} for the parameter $\kappa_{R}$. The plot for $\kappa_{R}$ has a strong frequency dependence and we find that $\kappa_{R}\left(\lambda_{1}\right)=1$, $\kappa_{R}\left(\lambda_{2}\right)=827$. Then, we obtain $\gamma_{32\left(31\right)}=\kappa_{R}\left(\lambda_{2}\right)\gamma^{\left(0\right)}_{32\left(31\right)}=8.27\cdot 10^{11} \; \textrm{s}^{-\textrm{1}}$ and $\gamma_{21}=\kappa_{R}\left(\lambda_{1}\right)\gamma^{\left(0\right)}_{21}=5\cdot 10^{8} \; \textrm{s}^{-\textrm{1}}$ ($\gamma^{\left(0\right)}_{21}=5\cdot 10^{8} \; \textrm{s}^{-\textrm{1}}$, $\gamma^{\left(0\right)}_{32}=\gamma^{\left(0\right)}_{31}=1\cdot 10^{9} \; \textrm{s}^{-\textrm{1}}$, see~\cite{Buckle}).

When the emitter is placed in a complex micro or nanostructured medium, the relaxation rate can be presented as
\begin{equation}
\label{eq:30}
\gamma_{ij}=\frac{\pi \omega_{ij}}{\hbar \varepsilon_{0}}\left|\mu_{ij}\right|^{2}\rho \left(\omega_{ij},\overline{r}\right),
\end{equation}
where $\mu_{ij}$ are the dipole moments of corresponding transitions. For vacuum we have $\rho \left(\omega_{ij},\overline{r}\right)=\frac{\omega^{2}_{ij}}{3\pi^{2}c^{3}}$ and consequently $\gamma^{\left(0\right)}_{ij}=\frac{\omega_{ij}}{3\pi \hbar \varepsilon_{0}c^{3}}\left|\mu_{ij}\right|^{2}$. In the case of an arbitrary medium, but for fixed orientation $\boldsymbol{\mathrm{u}}$ of the dipole, the equation for LDOS can be represented as
\begin{equation}
\label{eq:31}
\rho_{\boldsymbol{\mathrm{u}}}\left(\omega_{ij},\overline{r}\right)=\frac{2\omega}{\pi c^{2}}\textrm{Im}\left[\boldsymbol{\mathrm{u}}\boldsymbol{\mathrm{G}}^{E}
\left(\overline{r},\overline{r},\omega_{ij}\right)\boldsymbol{\mathrm{u}}\right],
\end{equation}
where $\boldsymbol{\mathrm{G}}^{E}\left(\overline{r},\overline{r},\omega_{ij}\right)$ is the electric Green function, $\overline{r}$ is the radius-vector of the NW position. Finally, in the case of $x$-oriented waveguide mode in Fig.~\ref{fig:7} we can present LDOS in the form
\begin{equation}
\label{eq:32}
\rho\left(\omega,\overline{r}\right)=\frac{1}{3\pi^{2}c}\frac{\left(n^{\left(\textrm{R}\right)}_{EF+}
\left(\omega\right)\right)^{2}}{\lambda^{2}_{0}}\varkappa \left(\omega,\overline{r}\right)^{2}
\end{equation}
owing to the reduction of the characteristic wavelength by a factor $n^{\left(\textrm{R}\right)}_{EF+}\left(\omega\right)$ and taking into account the spatial distribution of the field $\varkappa \left(\omega,\overline{r}\right)=\frac{E\left(\omega,\overline{r}\right)}{E^{\left(\textrm{max}\right)}\left(\omega\right)}$ in the waveguide, normalized to the maximum value $E^{\left(\textrm{max}\right)}\left(\omega\right)$. As a result, we have $\gamma_{ij}=\kappa_{R}\left(\omega\right)\gamma^{\left(0\right)}_{ij}$, where $\kappa_{R}\left(\omega\right)=\left(n^{\left(\textrm{R}\right)}_{EF+}\left(\omega\right)\varkappa \left(\omega,\overline{r}=\overline{r}_{c}\right)\right)^{2}$ for radius-vector $\overline{r}_{c}$ of the NW  center. Using the parameters from Tab.~\ref{tab:1} and Tab.~\ref{tab:3} and extracting information about $\varkappa \left(\omega,\overline{r}=\overline{r}_{c}\right)$ from full-wave simulation, we obtain $\gamma_{21}=1.013\cdot 10^{11} \; \textrm{s}^{-\textrm{1}}$, $\gamma_{32}=\gamma_{31}=1.094\cdot 10^{12} \; \textrm{s}^{-\textrm{1}}$ ($\gamma^{\left(0\right)}_{21}=5\cdot 10^{8} \; \textrm{s}^{-\textrm{1}}$, $\gamma^{\left(0\right)}_{31}=\gamma^{\left(0\right)}_{32}=1\cdot 10^{9} \; \textrm{s}^{-\textrm{1}}$, $\varkappa_{1}=0.1045$, $\varkappa_{2}=0.5577$). Note that the obtained result slightly differs from the previously obtained analytical results for an emitter near a flat graphene sheet.

We describe the energy of induced SPP-NW interaction using the coupling constants $g_{1\left(2\right)}\left(\overline{r}\right)=\sqrt{\frac{\omega_{1\left(2\right)}}{\hbar \varepsilon_{0}V_{EF1\left(2\right)}}} \varkappa_{1\left(2\right)}\left(\overline{r}\right)\mu_{12\left(32\right)}$, where $\mu_{12\left(32\right)}$ are the dipole moments of corresponding transitions in NW; $\varkappa_{1\left(2\right)}\left(\overline{r}\right)=\frac{E_{1\left(2\right)}\left(\overline{r}\right)}
{E^{\left(\textrm{max}\right)}_{1\left(2\right)}}$, $E_{1\left(2\right)}\left(\overline{r}\right)\equiv E\left(\omega_{1\left(2\right)},\overline{r}\right)$; $V_{EF1\left(2\right)}=\left(\frac{\lambda_{1\left(2\right)}}{n^{\left(\textrm{R}\right)}_{EF+}}\right)^{3}$ is the effective volume of interaction. Finally, we obtain $g_{1}=5.379\cdot 10^{12} \; \textrm{s}^{-\textrm{1}}$ and $g_{2}=3.318\cdot 10^{12} \; \textrm{s}^{-\textrm{1}}$.

\section{Tuning the parameters of pump SPP for switching the stub-resonator loaded with NW from the locking regime to the transmitting regime of signal SPP}
Our goal is to induce in a graphene waveguide both pump SPP at a wavelength $\lambda_{1}=2.56 \; \textrm{\textmu m}$ and signal SPP at a wavelength $\lambda_{2}=\lambda_{0}=8.04 \; \textrm{\textmu m}$ and to choose such $\Omega_{1}=g_{1}B$ and $\Omega_{2}=g_{2}a$ and frequency detunings to provide an additional phase shift of signal SPP $\Delta \phi_{\textrm{max}}$ equals to $\pi$ (shift on half wavelength). Here $a$ and $B$ are the amplitudes of signal and pump SPPs, respectively. The additional phase shift is given by $\Delta \phi_{\textrm{max}}=\frac{2\pi}{\lambda_{2}}n^{\left(\textrm{R}\right)}_{NW}D_{NW}$ and must be provided with a large value of correction to the refractive index $n^{\left(\textrm{R}\right)}_{NW}$ of NW material induced by strong nonlinear interaction between SPP modes and NW and described by (\ref{eq:24}), where $n_{NW}=n^{\left(\textrm{R}\right)}_{NW}+in^{\left(\textrm{I}\right)}_{NW}$. The correction to the complex refractive index can be expressed in the form $n_{NW}\approx {\chi_{NW}}/{2}$, where $\chi_{NW}=\frac{N\mu_{32}}{\varepsilon_{0}E_{2}}\overline{\rho}_{32}$ is the resonant part of the NW susceptibility, $N=5\cdot 10^{19} \; \textrm{cm}^{-\textrm{3}}$~\cite{Madelung} is the carrier concentration, $E_{2}$ is the signal field strength. Hence, we obtained the necessary value of the matrix element $\overline{\rho}_{32}$ to realize the required phase shift in the stationary regime for signal SPP, see (\ref{eq:24}b). It corresponds to $\textrm{Re}\left(\overline{\rho}_{32}\right)=0.0717$.
\begin{figure}[t]
\centering
\includegraphics[width=1.\columnwidth]{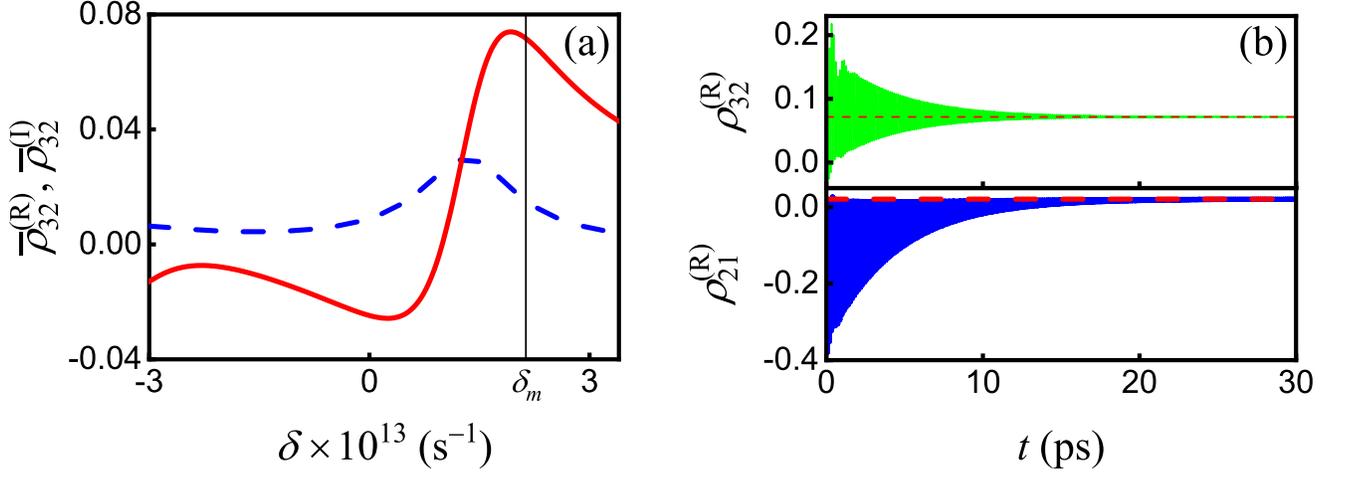}
\caption{\label{fig:10} (a) The frequency dependencies of real (solid red line) and imaginary (dashed blue line) parts of $\overline{\rho}_{32}$ for fixed detuning $\Delta=\Delta_{m}$; (b) the time dependencies of the real parts of $\rho_{32}$ (thin green and red lines) and $\rho_{12}$ (thick blue and red lines) calculated by using formulas (\ref{eq:24}) (dashed lines) and by using direct numerical simulation (solid lines) of the full system of differential Equations (\ref{eq:20}) for density matrix elements upon Ladder-type interaction of two SPP modes and core-shell NW.}
\end{figure}

We chose the amplitude of the signal field equals to $1$ photon ($a=1$), and the amplitude of the pump field equals to $4$ photons ($B=4$) and obtained $\Omega_{1}=2.151\cdot 10^{13} \; \textrm{s}^{-\textrm{1}}$ and $\Omega_{2}=3.318\cdot 10^{12} \; \textrm{s}^{-\textrm{1}}$ with an efficiency $E_{1\left(2\right)}\left(\overline{r}\right)=\varkappa_{1\left(2\right)}
\left(\overline{r}\right)E^{\left(\textrm{max}\right)}_{1\left(2\right)}$. Based on calculated Rabi frequencies $\Omega_{1\left(2\right)},$ relaxation rates $\gamma_{ij}$ and obtained stationary solutions (\ref{eq:24}), we plotted the frequency dependencies of the complex matrix element $\overline{\rho}_{32}$ and determined that the necessary value $\textrm{Re}\left(\overline{\rho}_{32}\right)=0.0717$ corresponds to parameter values $\Delta_{m}=-2\cdot 10^{13} \; \textrm{s}^{-\textrm{1}}$ and $\delta_{m}=2.132\cdot 10^{13} \; \textrm{s}^{-\textrm{1}}$ (see Fig.~\ref{fig:10}a). Further, we determined that obtained stationary solutions completely agree with the results of direct numerical simulation of the system (\ref{eq:20}) (see Fig.~\ref{fig:10}b). All calculated parameters are summarized in Table~\ref{tab:5}.
\begin{table}[H]
\caption{The stationary solutions of system (\ref{eq:20}) and corresponding frequency detunings.}
\label{tab:5}
\centering
\footnotesize
\begin{tabular}{*{9}{c}}
\toprule
$\Delta, \; \textrm{s}^{-\textrm{1}}$ & $\delta, \; \textrm{s}^{-\textrm{1}}$ & $\textrm{Re}\left(\overline{\rho}_{32}\right)$ & $\textrm{Im}\left(\overline{\rho}_{32}\right)$ & $\overline{\rho}_{11}$ & $\overline{\rho}_{22}$ & $\overline{\rho}_{33}$ & $\overline{n}_{21}$ & $\overline{n}_{32}$ \\
\midrule
$-2\cdot 10^{13}$ & $2.132\cdot 10^{13}$ & $0.0318$ & $0.0081$ & $0.4838$ & $0.4930$ & $0.0232$ & $0.0093$ & $-0.4699$ \\
\bottomrule
\multicolumn{9}{c}{} \\
\toprule
\multicolumn{2}{c}{$\overline{\rho}_{21}$} & \multicolumn{2}{c}{$\overline{\rho}_{32}$} & \multicolumn{2}{c}{$\overline{\rho}_{31}$} & & & \\
\midrule
\multicolumn{2}{c}{$0.0211+0.0035i$} & \multicolumn{2}{c}{$0.0717+0.0153i$} & \multicolumn{2}{c}{$-0.0668-0.0217i$} & & & \\
\bottomrule
\end{tabular}
\end{table}
\begin{figure}[t]
\centering
\includegraphics[width=0.43\columnwidth]{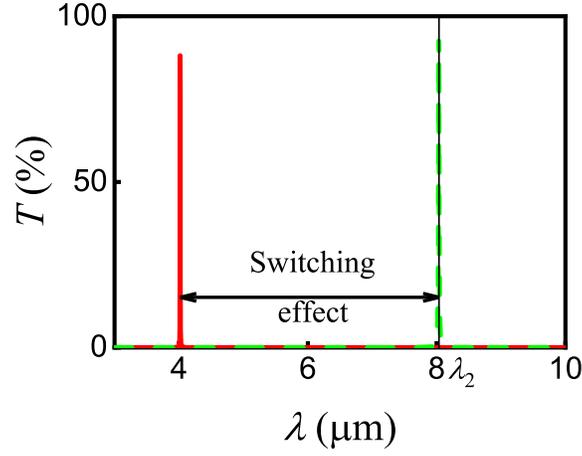}
\caption{\label{fig:11} The signal SPP transmittance for the stub nanoresonator with InAs/ZnS NW in the vicinity of $\lambda_{2}$ in the absence (solid red line) and in the presence of pump SPP mode $E_{1}$ (dashed green line).}
\end{figure}
\begin{figure}[b]
\centering
\includegraphics[width=0.5\columnwidth]{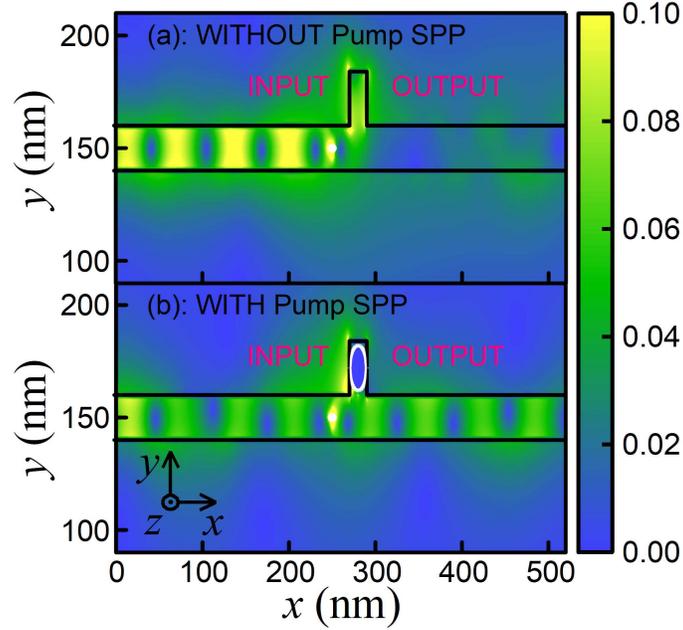}
\caption{\label{fig:12} The summarized electric field $\sqrt{E^{2}_{x}+E^{2}_{y}}$ distributions (arbitrary units) for signal SPP in the stub nanoresonator loaded with NW. The switching between regimes of (a) locking and (b) transmitting of signal SPP is demonstrated. The black lines correspond to the graphene waveguide with stub nanoresonator, and the circled white line depicts the NW.}
\end{figure}

Next, we calculated the transmittance (\ref{eq:16}) of the signal SPP near the wavelength $8.04 \; \textrm{\textmu m}$ for two cases, in the absence and in the presence of pump SPP, see Fig.~\ref{fig:11}. The appearance of pump SPP resulted in an additional phase shift $\Delta \phi_{\textrm{max}} =\pi$ contributes to the total phase shift of signal SPP $\phi \left(\lambda\right)=\frac{2\pi \left(2D+d\right)n^{\left(\textrm{R}\right)}_{EF+}}\lambda+\Delta \phi$. Under such conditions, we obtain the first-order constructive interference in stub nanoresonator, when $\Delta S=\lambda_{0}$. As you can see in Fig.~\ref{fig:11}, the presence of the required phase shift changes the transmittance of signal SPP from minimum to maximum values at $8.04 \; \textrm{\textmu m}$. The coefficients $r_{i}$, $t_{i}$, $s_{i}$ in this work, we chose empirically, in particular, $\left\{u\right\}=\left(0.1,0.9,0.065,0.9\right)$, where $\left\{u\right\}=\left(t_{1},s_{1},s_{3},r_{3}\right)$.

In order to verify the correctness of our analytical estimations, we carried out the direct numerical simulation taking into account the Ladder-type interaction of SPP modes with NW in the stub nanoresonator. We found the complete agreement of our numerical results with the theory when the presence of pump SPP leads to opening the transistor and switching to the regime of signal SPP transmitting, see Fig.~\ref{fig:12}. Finally, we estimated the switching time of the presented effect and it is about $20 \; \textrm{ps}$, which corresponds to a clock frequency of $50 \; \textrm{GHz}$. At the same time, during the process of switching, the transmittance increases from $7\%$ to $93\%$.

\begin{figure}[t]
\centering
\includegraphics[width=0.43\columnwidth]{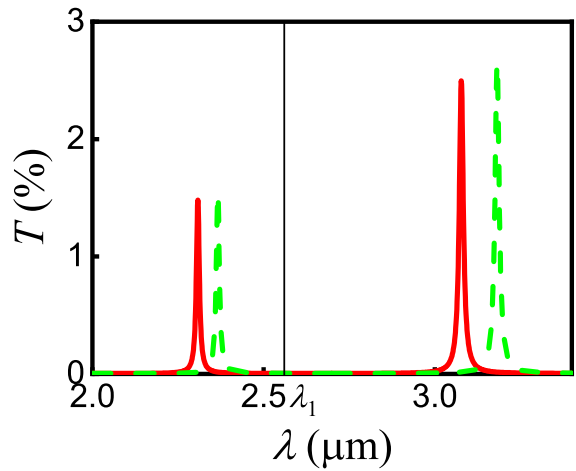}
\caption{\label{fig:13} The pump SPP transmittance for the stub nanoresonator in the vicinity of $\lambda_{1}$ in the absence (solid red line) and in the presence (dashed green line) of signal SPP.}
\end{figure}

Besides, in the process of interaction, the pump SPP also get additional phase shift $\Delta \phi_{\textrm{max}12}$ due to the arising of susceptibility $\chi_{NW12}=\frac{N\mu_{12}}{\varepsilon_{0}E_{1}}\overline{\rho}_{12}$ for NW. Under the selected conditions, this resonant shift is $\Delta \phi_{\textrm{max}12}=0.619=0.197\pi$ radians and the transmittance of the pump SPP will change in comparison with the empty stub, as in Fig.~\ref{fig:13}. Nevertheless, this change is not dramatically and the regime of pump SPP propagation is kept for stub loaded with NW.

\section{Conclusion}
The paper addresses the challenges of achieving a strong coupling regime in the process of interaction between graphene SPPs and semiconductor NW under the high-LDOS condition. We have proposed the model of all-plasmonic switcher based on a graphene stub nanoresonator loaded with core-shell NW and discussed the issues of creating such a device. It should be noted that the relatively short SPP propagation lengths in graphene systems, compared with MDS structures~\cite{Fedyanin}, significantly restrict now the scaling of such devices up to the level of integrated circuits~\cite{Ni}. At the same time, the presented model can be of fundamental importance for the development of both single high-speed switchers and devices based on them using hybrid metal-graphene structures~\cite{Yakubovsky,Chen2}.

Problems in the design and manufacture of all-plasmonic switchers require special attention. The creation of such devices is possible within the already available modern technologies, but using a combination of several different experimental techniques at once. We will briefly discuss the possibilities of the experimental realization of such devices. Initially, we assume that we have \ch{SiO2} substrate with recess corresponding to the further stub nanoresonator. Then, using plasma-enhanced chemical vapor deposition (PECVD) method~\cite{Chun} for deposition of graphene on \ch{SiO2} substrate, it is possible to form a single graphene layer on the top surface of the substrate. The next step is to load the core-shell NW into a stub nanoresonator. For this purpose, we propose to use the atomic force microscopy (AFM) nanomanipulation technique~\cite{Ratchford}. The atomic force microscopy allows to manipulate by a single semiconductor NW and to place it into stub with the required accuracy. The polymer buffer layer between graphene and conventional gate dielectrics can be used to improve the device characteristics~\cite{Kim1,Farmer}. Such polymer coating allows achieving high carrier mobility values of over $8000 \; \textrm{cm}^{\textrm{2}}/\left(\textrm{V}\cdot \textrm{s}\right)$ at room temperature~\cite{Kim1} for graphene field-effect transistors using, for example, \ch{Al2O3} as the top-gate dielectric. The next step is to coat the NW and graphene sheet with dielectric. For example, the atomic layer deposition (ALD) method can be used for the deposition of dielectric on graphene~\cite{Alles,Jeon,Nayfeh} or on a polymer buffer layer~\cite{Kang}. Moreover, there exists an alternative way of creating a dielectric layer on the graphene. The electron beam evaporation (EBE) method allows depositing \ch{SiO2} dielectric on the graphene surface~\cite{Hwang1}. Thus, using PECVD, AFM, and ALD or EBE methods one can completely produce the graphene ``transistor'' shown in Fig.~\ref{fig:7} with required device characteristics.
\\

\section*{Author Contributions}
Conceptualization and methodology, A.V.P.; formal analysis, A.V.P., M.Yu.G. and A.V.S.; software and investigation, A.Yu.L., M.Yu.G. and A.V.P.; visualization, M.Yu.G. and A.V.S.; writing-original draft, A.V.P. and M.Yu.G.; writing-review and editing, V.S.V. and A.Yu.L.; conceptualization and funding acquisition, V.S.V.

\section*{Conflicts of Interest}
The authors declare no conflict of interest

\begin{acknowledgments}
Authors acknowledge the Ministry of Science and Higher Education of the Russian Federation [grant number 0714-2020-0002].
\end{acknowledgments}

\bibliography{SwitchArXivRef}

\end{document}